\documentclass{article}
\usepackage[a4paper, total={6.5in,9in}]{geometry}
\usepackage{graphicx}
\usepackage{epstopdf}% To incorporate .eps illustrations using PDFLaTeX, etc.
\usepackage{subfigure}% Support for small, `sub' figures and tables
\usepackage{url}
\usepackage{authblk}
\usepackage{amsmath}
\usepackage{rotating}
\usepackage{amsfonts}
\usepackage{natbib}% Citation support using natbib.sty
\begin{document}

\vspace{-40pt}
\title{Reaction Asymmetries to Social Responsibility Index Recomposition: A Matching Portfolio Approach} 
\vspace{-30pt}

%\author[1]{Jari Kappi \thanks{Email: jari.kappi@xjtlu.edu.cn}}
\author[1]{Wanling Rudkin \thanks{Corresponding author. Tel: +44 7514023738 Email: wanling.qiu@liverpool.ac.uk}}

\author[1]{Charlie X. Cai} 
\affil[1]{Management School, University of Liverpool, Liverpool, United Kingdom}
\vspace{-20pt}
\maketitle
\vspace{-35pt}

\begin{abstract}
Listing on the Dow Jones Sustainability Index is seen as a gold-standard, verifying to the market that a firm is fully engaged with a corporate social responsibility agenda. Robustly quantifying the impact of listing, and de-listing, against any industry level shocks, as well as evolution in the competitive relationship between firms within the industry, provides a strength absent in existing works. It is shown that cumulative abnormal returns on stocks added to the index are significantly positive in the three trading weeks prior to the official announcement. The post-listing correction result posited to date is also demonstrated to hold; the proportion of periods with significant negative returns is low, however. Announcement, rather than effective dates are critical to returns. Differentials between these stages in the chronology is an important contribution of this paper. Most effects end before the membership changes become effective. Whilst there are considerable gains to be made, they come pre-announcement date and require foresight to exploit. Investors must research likely new members to gain maximum return. 
\end{abstract}


\maketitle

\section{Introduction}

%Corporate social responsibility (CSR) has been demonstrated to impact corporate financial performance; typically it is assumed that the outcome stems from consumer preference  for goods from CSR practitioners outweighing costs of implementing responsible practice. Improved profitability, and hence stronger future cash flows, have been disputed, but those who obtain higher CSR recognition are found to have lower volatility, positing an argument for lower returns. Evaluating the balance between stronger future cash flows and reduced volatility, this paper asks whether there are any positive impacts on stock returns and, if so, where they occur relative to the public announcement day. Inclusion on the DJSI is binary, firms are on or they are not, making binary models the start point for any listing study. Generating results from two-sample comparisons and dummy variable regression approaches we verify that our dataset produces the same ambiguity of conclusions that characterise other studies on CSR and returns. Insights developed in what follows have demonstrable robustness to industry, or economy wide level shocks; the reference for abnormal returns is not the past performance of the share but the present performance of a portfolio of shares weighted to match the performance of the share over a long pre-treatment period. This paper thus contributes deeper understanding of the impact of CSR behaviour recognition on stock returns in a framework attentive to wider conditions, competition and the relative performance within the sector. 

Increasing recognition of the importance of corporate social responsibility (CSR) to firms performance is drawing natural attention to the subsequent link to stock returns. An implicit trade off between enhancing reputation and the costs of implementing improved practices has troubled analysts reconciling the observed trends of firms striving to achieve market leadership on CSR. Many theories have been posited for the effect sustainable practice will have on stock returns, resulting in an ambiguity about what response will be observed. Isolating a CSR effect on stock returns requires three key elements to ensure correct identification. Firstly there must be agreed empirical evidence that a firm is practising CSR, this must be independently verifiable to be trustworthy to investors. Secondly, information about a firms CSR leadership must be visible to the market through a clearly identifiable channel. For \cite{fowler2007critical} these must be verified by an assessor who is independent from the firm. Finally, it must be possible to isolate decisions based upon that information from all other characteristics of the firm that might otherwise influence stock returns. This paper meets the challenge using the Dow Jones Sustainablity Index (DJSI), which sees firms independently assessed for inclusion and evidences sustainability in a binary fashion. That listings are reappraised only once each year provides an ideal test point for the impact of CSR leadership on stock returns.    

Generating results from two-sample comparisons and dummy variable regression approaches we verify that our dataset produces the same ambiguity of conclusions that characterise other studies on CSR and returns. Insights developed in what follows have demonstrable robustness to industry, or economy wide level shocks; the reference for abnormal returns is not the past performance of the share but the present performance of a portfolio of shares weighted to match the performance of the share over a long pre-treatment period. This paper thus contributes deeper understanding of the impact of CSR behaviour recognition on stock returns in a framework attentive to wider conditions, competition and the relative performance within the sector.

Begun with a global index in 1999 with a global index, the DJSI family has expanded rapidly to include a number of regional and themed indices. First of the expansions was the creation of North American list to cover the United States and Canada; this is the index that is referred to as DJSI in all that follows. In each case the principle of determining which firms will be listed is highly similar, the leading firms in each industry being selected. Early establishment of the DJSI as a measure of CSR leadership is provided in the works of \cite{fowler2007critical} and \cite{hawn2018investors}. For all indexes firms are invited to submit a wealth of documentation for evaluation, this weighty evidence then informing the inclusion decision of Robecco Sustainable Asset Management (Robecco SAM). Assessment of firms is constantly evolving to reflect current best practice, helping maintain the association between listing and market leadership. Leadership is questioned by \cite{ziegler2010determines} and \cite{oberndorfer2013does}, who contend that the signal is actually one which says that firms are effectively following competitive strategies that respond to the market for pecuniary advantage. Responses to this critique note that in innovating to pursue such strategies firms do become leaders. 

Although there is disagreement on measuring CSR, a listing on one of the DJSI indexes signals clearly to the market that a firm is meeting critical Corporate Social Responsibility (CSR) standards\footnote{Disagreement about CSR measurement is charted in \cite{scalet2010csr} and subsequently \cite{venturelli2017can}.}. As a measure of CSR it is binary, making interpretation simple. Firms are clear leaders in their market, or they are not; membership of the DJSI identifying the former. Studying DJSI listing, or continued membership, becomes simpler than discerning the marginal effects of indexes that are reported for only subsets of the wider set of traded firms\footnote{Continuous measures are often born from research carried out by teams at large agencies such as MSCI KLD. From here emerges either a scale reading or a series of binary evaluations of strengths and concerns that then form the measure using net strengths. Whilst not continuous there are advantages over the low level of splitting offered by the DJSI dummy. However, as \cite{mattingly2017corporate} notes in reviewing the dataset there are challenges of subjectivity in measure construction and problems of data coverage outside the biggest firms.}. Such binary delineation also facilitates event studies and treatment effect approaches of the type performed here. From an investor behaviour perspective it is implicitly assumed that listing conveys information more effectively than discerning marginal effects amongst metrics of CSR activity. Evidence on the benefit of simple communication is provided by \cite{hartzmark2019investors} review of investor response to the Morningstar globe ratings for mutual fund sustainability\footnote{The Morningstar globes rate the sustainability of the holdings of all mutual funds listed on Morningstar, one of the world leading fund websites. Funds are given a rating between one and five globes, with the latter informing on leadership.}. Information used for their construction is all public meaning the large reaction demonstrated may be entirely attributed to the simplicity of the rating.

Event studies have advantage where timings are known and exogenous to the units being considered \cite{mackinlay1997event}. Listing on a social responsibility index, such as the DJSI, is completely exogenous from the share price of a particular firm. Likewise although the inclusion of a firm into the index is a result of the firms efforts it is timed at a point dictated by the listing agency. It is this that offers the requisite exogeneity. The financial literature expanding upon CSR events focuses on either company specific events, or exogenous occurrences that impact a subset of stocks\footnote{\cite{clacher2012announcements} and \cite{cai2014corporate} fall into the category of single company events}. Index listing, or de-listing, fits firmly into this second class\footnote{Wider consideration of listing as an event drives \cite{denis2003s} evaluation of learning expectations following inclusion in the S\&P 500 index. A large literature on listing effects follows in this mould.}. An endogeneity in listings based on financial performance may be argued, the direction of the relationship being blurred in empirical work \citep{scholtens2008note}. However, for social indexes where assessments are undertaken, and decisions made, far ahead of the announcement there will be no changes close to listing dates that would provide any information about the likelihood of listings.

Evaluations of the effect of joining the DJSI have applied event studies on listing announcements \citep{cheung2011stock,robinson2011signaling,lourencco2014value,joshi2017asymmetry,hawn2018investors}. Consistent through all of these works is the belief that there needs to be consideration of the time before, and after, the announcement rather than a focus purely on the announcement week itself. \cite{hawn2018investors} does not go beyond a few days but the data in their tables suggests there may be effects across longer date ranges. Motivation for longer time frames comes from the imperfect information in the markets and the ability of some traders to form meaningful expectations of any upcoming listing. There is evidence of a number of abnormal movements presented here that is consistent with information being available to only some investors. Evaluation of the abnormal returns therefore begins three weeks before the actual announcement, with most pre-announcement changes occurring around two weeks ahead of the official release of the DJSI constituent list. A further consistency lies in the creation of abnormal returns in the short term, but that all shares revert to their expected levels within a few weeks of the announcements\footnote{International studies likewise find evidence of short-term effects \citep{oberndorfer2013does,orsato2015sustainability,nakai2013sustainability}.}.

\cite{oberndorfer2013does} posit two competing hypotheses for the short term impact of DJSI listing, a revisionist and a traditionalist perspective. Revisionists argue that inclusion represents a commitment to stakeholders that boosts sales, increases employee happiness and hence productivity, is hard for rivals to compete against and sets an upward trend of financial performance. Consequently the revisionist argument is summarised by \cite{oberndorfer2013does} as leading to positive short term listing effects. Contrary to this the traditionalist approach contends that focus on CSR diverts resources from productive endeavours, reducing productivity and hence lowering profitability. In practice investors may expect either one of these hypotheses to hold true. A signal of sustainability leadership from a DJSI listing may thus result in either an increased, or decreased, return dependent on the dominant effect. Results on listings presented herein, consistent with the pre-announcement and correction effects, point to the revisionist angle dominating those receiving news ahead of the listing. In this paper that is also true for the immediate post listing announcement period also. Corrections may suggest that the traditionalist effect dominates as the changes become effective, but equally may simply signal that the market believes the original revisionist appraisal was too optimistic. In the case of firms which exit the list the traditionalist approach drives positive returns, with the subsequent correction being in the revisionist direction. In both cases the hype around the listing date is driving much larger effects than the market ultimately displays, the correction effect bringing the models back towards their past trends.

Other studies consider alternative indices such as the Newsweek Green Rankings\footnote{The Newsweek Green Rankings were first released in 2009 and gained wide interest in the USA. Scores are constructed as a combination of an environmental impact score (45\%) using emissions data, green policies (45\%) which are obtained in part from the KLD database, and a green reputation score (10\%) based on a survey of relevant stakeholders and academics \citep{cordeiro2015firm}. These are thus more environmentally focused than the DJSI index which captures more of the social responsibility range.}  \citep{cordeiro2015firm} and the World's Most Ethical Companies list \citep{karim2016ethical}. As public facing measures these have garnered greater media coverage. \cite{cordeiro2015firm} hypothesised higher rankings in the 2009 listing would correspond to stronger favourable reactions, and that this would continue both short-term and long-term; evidence of such effects is found. For works dependent on such single-year orderings there is an inevitable problem of repeatability; we demonstrate subsequently the impact of DJSI listing is significantly different during the period studied by \cite{cordeiro2015firm}.

Methodologically, synthetic control approaches after \cite{abadie2010synthetic}, offer natural synergies in finance, where their allocations of weightings to a series of assets to create a portfolio that recreates the asset of interest is synonymous with exchange traded funds. This motivates work on the effect of political connections to the Trump administration following the 2016 presidential election \citep{acemoglu2016value}, the impact of the Arab Spring on Egyptian markets relative to others \citep{acemoglu2017power}, as well as \cite{chamon2017fx} work on currency interventions in Brazil. In each case comparison with a portfolio of assets is championed as effective in capturing the change from the treatment, be that political connections, the position of the Egyptian market, or the Brazilian currency. This paper preserves the benefits exposited in these papers, whilst simultaneously introducing an ability to cope with multiple listings from the same asset group in the same period. 

Whilst the \cite{abadie2010synthetic} approach has much to offer, an inability to handle multiple treatment units simultaneously, and computationally intensive confidence interval calculation, have limited adoption. This paper responds using the generalised synthetic control \citep{xu2017generalized} \footnote{This development from synthetic to generalised synthetic has been kept up with in the political economy literature. Interested readers are direct to \cite{abadie2015comparative}.}. Primary advantages of so doing are the ability to produce treatments that recognised the simultaneity derived from a single potential listing date per year. Further the software implementation of the generalised synthetic control, \textit{gsynth} \citep{gsynth}, generates all necessary confidence intervals meaning there is no need to manually implement the bootstrapping approach of placebo treatments used in \cite{acemoglu2016value}. 

Three key contributions are made to the literature. A primary contribution comes in the formalisation of the effect of pre-announcement leakage of news about changes in firm's DJSI membership status. We evidence some post-change correction on those firms which join the DJSI and those who are announced as being delisted. Such is as rationality may expect. We demonstrate that the effects occur further from the effective date than suggested by simple construction of abnormal returns against a CAPM fit. Quantitatively different investment direction is thus found. In recognising simultaneous treatments we show the market also moves post announcement reducing the extent to which any increased returns can be classed as abnormal. Through its treatment of multiple firm listings at the same time this paper meets the challenges laid down in the recent work on the synthetic control method by \cite{abadie2015comparative,acemoglu2016value} and others. Thirdly, the recognition of contemporaneous treatment developed here can be readily ported out of these listing effects studies and into the wider event study framework. Across these three contributions we deliver a deeper, more robust, evaluation of the effect of listing on the DJSI upon abnormal stock returns.

The remainder of the paper is organised as follows. Data, abnormal return construction and financial controls are introduced in Section \ref{sec:data}. Two-sample results, and OLS regressions that measure DJSI listing as a dummy variable are set out in Section \ref{sec:first} as a comparator for the subsequent analysis. Section \ref{sec:synth} details the generalised synthetic control method and gives results from the comparisons between listed shares and counterfactual alternatives constructed on the assumption those firms did not join the DJSI. Section \ref{sec:discuss} reviews the information gained from this robust analysis before Section \ref{sec:conclude} reinforces the value of the work and the ways in which useful extensions may be made.

\section{Data and Empirical Approach}
\label{sec:data}

When considering changes in the constituents of the DJSI as a result of the annual review there are four possible combinations of before, and after, status. Firstly there are those firms who are listed on the DJSI and, following the announcement are still listed. These are the firms who stay on the list and continue to be recognised for their CSR leadership. Likewise there are those firms for whom standards did not, and still do not, meet DJSI inclusion criteria, these are the firms who stay off and again see no change in their DJSI status. In this paper focus is on those firms who change status. Those who meet the assessment criteria will gain listing, whilst those whose standards fall will lose their place. 

Recognising industry differentials in the assessment criteria, and recognising unobserved heterogeneity between industries in stock returns, all control samples for the assessment of listing status changes are drawn from the same industry. Given that not all industries will have a listed, or de-listed firm, in a given announcement it follows that the samples used to study the listing and de-listing effects will differ. This and subsequent sections run the analysis of listing and de-listing contemporaneously to ease comparisons. 

A key delineation is made between comparisons run on the full set of control firms and those which first eliminate a subset of the potential controls to create balance. This full sample versus base sample approach is replicated here for similar motivations. Details of the sample construction process follow. Primary evaluation of listing, or de-listing effects is made using abnormal returns on a daily basis. These returns may be combined over time to produce a mean square percentage error as a barometer of model fit. That construction is also detailed below. Finally we present the descriptive statistics incorporating listing/de-listing and full/base samples. 

\subsection{Full Sample}
\label{sec:full}

Data on constituents of the DJSI is constructed using listings from RobeccoSam, with entries recorded for each year\footnote{An introduction to the index, containing details of how to construct an entrants list, is provided in \cite{robecosam2013dow}.}. For each listing, or de-listing, the North American Industry Classification System (NAICS) code is obtained at the two-digit level. Utilising unambiguous industry definitions in this way facilitates the formation of a control sample from the same industry, guarding against impacts from industry level heterogeneities. Share price data comes from CRSP and is gathered daily for the period beginning the first of November in the year prior to the listing, to a date 15 days after the listing becomes effective. This results in up to 250 observations for each firm. In order to be included in the samples the firms must have sufficient numbers of observations throughout the studied period. Data on firms accounting fundamentals is taken from Compustat. Data is merged such that the accounting data from the financial year previous to the announcement being studied is used as controls. Given all announcements take place in September such an alignment approach is consistent with the established practice of using firm financial characteristics for the previous calendar year only after July 31st.

\begin{figure}
	\begin{center}
		\caption{Listing Timeline \label{fig:time}}
		\includegraphics[width=14cm]{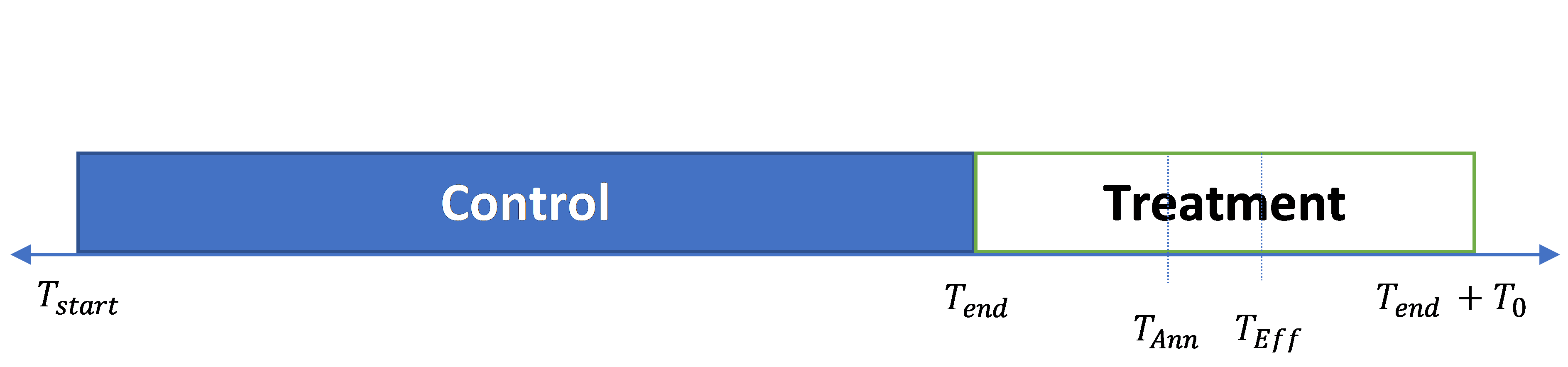}
	\end{center}
\raggedright
\footnotesize{Notes: Control refers to the period over which models are trained, beginning at time $T_{start}$ and ending 16 days prior to the announcement at time $T_{end}$. The length of the control period is defined as $T_c$ and represents the difference in trading days between $T_{start}$ and $T_{end}$ The subsequent day,$T_{end}+1$, is the first in the treatment period over which models are assessed. This treatment period ends after $T_0$ periods at time $T_{end}+T_0$. Within the treatment period there are two key dates $T_{Ann}$ when the announcement of changes to the constituents is made and $T_{Eff}$, one week later, when those changes become effective. Announcement periods vary by year, but in all cases $T_{start}$ is the 1st November in the year prior to the announcement being studied.}
\end{figure}

Figure \ref{fig:time} depicts the periods discussed in the exposition that follows. In all cases the specific time (trading day) being considered is referred to as $t$. A control period, $t \in \left[T_{start},T_{end}\right]$ is defined as the period over which all models are trained. Model performance is then evaluated in the treatment period which begins on day $T_{end}+1$ and ends on day $T_{end}+T_0$, $T_0$ days later. Following past works the treatment period extends 15 trading days before the index composition announcement and ends 15 days after the effective date. This represents a change from many studies who only base their period around as single treatment day. There are 5 trading days, 1 week, between the announcement and effective dates such that the total period is 36 days. Henceforth we can think of the treatment period as capturing $t \in \left[1,36\right]$ as the time frame for which we are evaluating listing/de-listing effects, with $T_0=36$. In this way announcement day becomes day 16 and the effective day is day 21.  Consequently, the treatment period extends through October, ending on a different date each time. To ensure that there is no overlap, and that there is no impact of any past membership announcements on subsequent periods, the control period for the subsequent year does not begin until 1st November. This pattern repeats for each year between the formation of the DJSI North America in 2005 and 2018, being the final year for which we have the data completed. Our share data thus runs from 1st November 2004 to 16th October 2018, 15 trading days after the 24th September 2018 effective date. De-listing requires that a firm first be on the DJSI North America; the first delistings are only found in the 2006 announcements.

\begin{table}
	\begin{center}
		\begin{tiny}
			\caption{Treatment and Control Numbers \label{tab:tandc1}}

			    \begin{tabular}{llrrrrrrrrrrrrrrrrrr}
			    \hline
			    & & \multicolumn{18}{c}{NAICS 2-Digit Industry Code} \\
				&& 21     & 22    & 23    & 31    & 32    & 33    & 42    & 44    & 45    & 48    & 51    & 52    & 53    & 54    & 56    & 62    & 72    & \multicolumn{1}{l}{Total} \\
				\hline
				\multicolumn{20}{l}{Panel A: Listings}\\
				2005&L     & 1     & 6     & 1     & 3     & 8     & 9     &       & 4     & 1     & 1     & 4     & 7     &       & 2     & 1     &       & 1     & 49 \\
				&C     & 129   & 99    & 31    & 105   & 318   & 661   &       & 89    & 53    & 89    & 285   & 535   &       & 112   & 57    &       & 62    & 2625 \\
				2006&L     & 2     & 4     &       & 1     & 3     & 1     & 1     &       &       &       & 1     & 3     &       &       &       &       &       & 16 \\
				&C     & 140   & 104   &       & 109   & 327   & 683   & 82    &       &       &       & 301   & 540   &       &       &       &       &       & 2286 \\
				2007&L     & 1     & 1     &       &       &       & 2     &       &       & 1     &       & 1     & 3     &       & 1     &       & 1     &       & 11 \\
				&C     & 162   & 106   &       &       &       & 694   &       &       & 49    &       & 318   & 517   &       & 114   &       & 52    &       & 2012 \\
				2008&L     & 1     & 2     &       & 1     & 1     & 2     &       &       & 1     &       & 1     & 2     & 2     &       & 1     &       &       & 14 \\
				&C     & 145   & 100   &       & 104   & 282   & 573   &       &       & 34    &       & 256   & 440   & 156   &       & 51    &       &       & 2141 \\
				2009&L     & 1     & 2     &       & 1     & 4     & 2     & 2     & 1     & 1     & 1     & 1     & 3     &       & 1     &       &       &       & 20 \\
				&C     & 119   & 100   &       & 88    & 221   & 430   & 60    & 60    & 26    & 84    & 204   & 353   &       & 87    &       &       &       & 1832 \\
				2010&L     & 3     & 2     &       &       & 4     & 3     &       &       &       &       & 1     &       &       & 2     & 1     &       &       & 16 \\
				&C     & 141   & 106   &       &       & 274   & 572   &       &       &       &       & 248   &       &       & 111   & 46    &       &       & 1498 \\
				2011&L     &       & 1     &       & 1     & 1     & 4     &       & 1     & 1     & 2     &       & 3     & 1     &       & 1     &       & 1     & 17 \\
				&C     &       & 105   &       & 104   & 273   & 577   &       & 80    & 33    & 88    &       & 462   & 178   &       & 49    &       & 47    & 1996 \\
				2012&L     &       & 1     &       & 2     & 2     & 2     &       &       & 1     &       & 3     & 1     &       & 1     & 1     &       &       & 14 \\
				&C     &       & 98    &       & 107   & 275   & 574   &       &       & 36    &       & 263   & 499   &       & 107   & 48    &       &       & 2007 \\
				2013&L     & 2     &       &       & 2     & 4     & 4     &       & 2     &       &       & 3     & 2     & 3     & 1     &       &       &       & 23 \\
				&C     & 152   &       &       & 112   & 293   & 587   &       & 83    &       &       & 287   & 542   & 191   & 103   &       &       &       & 2350 \\
				2014&L     &       & 2     & 2     &       & 1     & 3     &       &       &       & 1     & 2     & 2     & 1     &       &       &       & 1     & 15 \\
				&C     &       & 100   & 46    &       & 349   & 637   &       &       &       & 119   & 322   & 591   & 220   &       &       &       & 58    & 2442 \\
				2015&L     &       &       & 1     &       & 3     & 1     &       &       &       &       & 1     & 1     & 2     &       &       &       & 1     & 10 \\
				&C     &       &       & 43    &       & 392   & 599   &       &       &       &       & 333   & 392   & 217   &       &       &       & 61    & 2037 \\
				2016&L     &       &       &       & 2     & 1     & 2     &       & 1     &       &       & 1     & 3     & 2     &       & 1     &       &       & 13 \\
				&C     &       &       &       & 105   & 381   & 560   &       & 81    &       &       & 329   & 564   & 223   &       & 53    &       &       & 2296 \\
				2017&L     & 2     & 1     &       & 2     & 2     & 3     & 1     &       &       & 1     & 2     & 2     & 2     & 1     &       &       & 2     & 21 \\
				&C     & 121   & 99    &       & 106   & 388   & 576   & 95    &       &       & 115   & 330   & 608   & 224   & 88    &       &       & 62    & 2812 \\
				2018&L     & 1     &       &       &       & 2     & 2     &       &       &       &       &       & 2     &       &       & 1     &       &       & 8 \\
				&C     & 103   &       &       &       & 221   & 401   &       &       &       &       &       & 429   &       &       & 28    &       &       & 1182 \\
				&&&&&&&&&&&&&&&&&&&\\
				\multicolumn{20}{l}{Panel B: De-listings}\\
				2006  & D     &       & 2     &       & 1     &       & 2     &       &       &       &       & 2     & 2     &       &       &       &       &       & 9 \\
				& C     &       & 106   &       & 109   &       & 682   &       &       &       &       & 300   & 541   &       &       &       &       &       & 1738 \\
				2007  & D     &       &       &       &       &       &       &       &       &       &       & 1     & 2     &       &       &       &       &       & 3 \\
				& C     &       &       &       &       &       &       &       &       &       &       & 318   & 518   &       &       &       &       &       & 836 \\
				2008  & D     &       &       &       &       & 1     & 1     &       &       &       &       & 1     &       &       &       &       &       &       & 3 \\
				& C     &       &       &       &       & 282   & 574   &       &       &       &       & 256   &       &       &       &       &       &       & 1112 \\
				2009  & D     &       &       &       &       & 1     & 1     &       &       &       &       &       & 1     &       &       &       &       &       & 3 \\
				& C     &       &       &       &       & 224   & 431   &       &       &       &       &       & 355   &       &       &       &       &       & 1010 \\
				2010  & D     &       &       &       &       & 4     & 2     &       &       &       &       & 1     &       &       &       & 1     &       &       & 8 \\
				& C     &       &       &       &       & 274   & 573   &       &       &       &       & 248   &       &       &       & 46    &       &       & 1141 \\
				2011  & D     &       &       &       &       &       &       &       &       &       &       &       & 1     &       &       & 1     &       &       & 2 \\
				& C     &       &       &       &       &       &       &       &       &       &       &       & 464   &       &       & 49    &       &       & 513 \\
				2012  & D     &       & 2     &       &       & 2     & 4     &       &       & 2     &       & 2     & 1     &       & 2     &       &       &       & 15 \\
				& C     &       & 97    &       &       & 275   & 573   &       &       & 35    &       & 264   & 499   &       & 106   &       &       &       & 1849 \\
				2013  & D     &       &       &       & 1     & 5     & 3     &       & 1     &       &       &       & 3     &       &       &       &       &       & 13 \\
				& C     &       &       &       & 113   & 292   & 588   &       & 84    &       &       &       & 541   &       &       &       &       &       & 1618 \\
				2014  & D     &       &       &       &       & 1     &       &       &       &       &       & 1     &       &       &       &       &       & 1     & 3 \\
				& C     &       &       &       &       & 349   &       &       &       &       &       & 323   &       &       &       &       &       & 58    & 730 \\
				2015  & D     &       &       & 1     & 1     &       & 4     &       &       &       &       &       & 1     & 1     &       &       &       &       & 8 \\
				& C     &       &       & 43    & 106   &       & 596   &       &       &       &       &       & 592   & 218   &       &       &       &       & 1555 \\
				2016  & D     &       &       &       & 1     & 1     & 2     &       &       &       &       &       & 1     &       &       & 1     &       &       & 6 \\
				& C     &       &       &       & 106   & 381   & 560   &       &       &       &       &       & 516   &       &       & 53    &       &       & 1616 \\
				2017  & D     & 3     &       &       & 2     & 1     & 1     &       &       &       &       & 1     & 1     & 1     &       &       &       &       & 10 \\
				& C     & 120   &       &       & 106   & 389   & 578   &       &       &       &       & 331   & 609   & 225   &       &       &       &       & 2358 \\
				2018  & D     &       &       &       & 2     & 2     &       &       &       &       &       &       & 2     &       &       &       &       &       & 6 \\
				& C     &       &       &       & 57    & 221   &       &       &       &       &       &       & 429   &       &       &       &       &       & 707 \\
				
				\hline
			\end{tabular}%
		\end{tiny}
	\end{center}
\raggedright
\footnotesize{Notes: Numbers represent the number of firms included in the full sample for the estimation of cumulative abnormal returns. L is used in Panel A to denote the number of firms joining the DJSI in the given year, with D used to denote the number of de-listings in Panel B. In both panels C denotes the numbers of controls. Totals are provided for each year. 2005 has more joining firms because this was the year that the DJSI North America was formed and numbers of listed firms in North America thus increased to populate the regional list. Numbers reflect those industry-years for which there is sufficient share price data, and for which assets are listed for the preceding financial year.}
\end{table}

In any given year the number of treated observations can vary, and many industries will not feature amongst either the newly listed set, or the de-listed set. Two digit NAICS codes, and the number of entering firms there from, are reported in Table \ref{tab:tandc1}. Panel A reports the numbers for listings (L), whilst Panel B provides numbers for de-listings (D). In each case the number of control firms is given by C. For the univariate and regression approaches these numbers do not present a challenge, but for the generation of counterfactual versions of listed shares the numbers in $L$, or $D$ are important. The original synthetic control method of \cite{abadie2010synthetic} allows for only one treated unit but it is clear from the numbers that many year-industry pairs have more than one entrant or exiting firm. In this case we can not ignore the potential impact that the other newly listed firm might have upon any other firm gaining DJSI listed status. Likewise effects of other de-listed firms also require control. Hence there is a call for a methodology that is robust to such diversity of treated unit profiles; \cite{xu2017generalized} employed in Section \ref{sec:results} meets this call.  

\subsection{Reduced Sample}
\label{sec:redsamp}

Amongst the full sample are a number of firms who are significantly smaller than any of those who are members of the DJSI. This creates a potential bias in the comparison due to the well studied size anomaly\footnote{See \cite{keim1983size} for a review of the work that established this anomaly within the asset pricing literature.}. Consequently a further control is placed upon firms that ensures the control set is more directly comparable with the treated set. Here a reduced sample is constructed using only those firms who have assets of at least 80\% of those of the smallest firm that joins the DJSI in that year. By imposing this restriction we significantly reduce the number of shares available to serve as comparators, but are able to minimise the impact of size. Alternative thresholds could be considered, but with the contribution of this paper stemming from an approach that does not require sample size reduction robustness of the results in Section \ref{sec:first} to minimum size is taken as given from the papers advocating those approaches. 

In the discussion of established modelling methodologies we present both the base sample and full sample, but do not use the base sample for the generalised synthetic control approach.

\subsection{Cumulative Abnormal Returns}
\label{sec:simpcar}

Evaluation of the effect of changes in a firm's DJSI listing status is based upon the ability of membership to generate returns which differ from those that might have been expected in the event that the firm did not receive the listing. This may be achieved either by comparing new entrants with similar firms that are not joining the DJSI that year, or by comparing de-listing firms with others who are not exiting the DJSI that year. However, it is more usefully considered as the difference between the observed returns and those that would have been realised had pricing behaviour of the listed firms share continued in the same way as it had been doing during the control period.

Simplest of the models to study the cross section of stock returns is the capital asset pricing model (CAPM) as introduced through the works of \cite{lintner1965security,sharpe1964capital} and \cite{treynor1962jack}. Although subsequent advancements of the CAPM are able to generate better fit for future returns predictions it is widely accepted that the CAPM is the most parsimonious solution for out-of-sample prediction [**** ADD CAMPBELL] \citep{acemoglu2016value}. Before proceeding note that in all that follows we could add an additional $y$ subscript to recognise that all estimation and prediction applies to a specific year and that there are multiple years in the dataset. For the control period, $t \in \left[T_{start},T_{end}\right]$, we estimate equation \eqref{eq:capm1} using ordinary least squares (OLS) regression. This is done for all firms in the sample individually. 
\begin{align}
R_{it}=\alpha_i + \beta_iMKT_t \label{eq:capm1}
\end{align}
In equation \eqref{eq:capm1} $R_{it}$ is the excess return on share $i$ at time $t$, $MKT_t$ is the Fama-French excess return for the market at time $t$, and $\alpha_i$ and $\beta_i$ are the coefficients of interest. Estimated values $\hat{\alpha}_i$ and $\hat{\beta}_i$ are then used to compute the fitted excess returns for share $i$, $\hat{R}_{it}$. The abnormal return, $AR_{it}$, is then defined as the difference between fitted and observed values:
\begin{align}
AR_{it}=R_{it}-\hat{\alpha}_i-\hat{\beta}_iMKT_t
\end{align}
Consequently a subperiod $t \in \left[\underline{t},\bar{t}\right]$ has cumulative abnormal returns (CAR)s of:
\begin{align}
CAR_i[\underline{t},\bar{t}]=\sum_{t=\underline{t}}^{\bar{t}}AR_{it} \label{eq:car2}
\end{align}
Investors have natural interest in obtaining abnormal returns, with higher absolute values being most attention grabbing. If correctly priced the CAR would be zero and hence the relationship between CARs and DJSI status becomes of interest.

This paper contrasts these simple abnormal returns with those generated by the synthetic control family. For this purpose we employ the mean square predicted error (MSPE) within the control period as a measure of model fit. For any given share $i$ the MSPE over the $T_c$ trading day interval $[T_{start},T_{end}]$ is given by equation \eqref{eq:mspe1}.
\begin{align}
MSPE_i=\dfrac{1}{T_c}\sum_{t=T_{start}}^{T_{end}}AR_{it}^2 \label{eq:mspe1}
\end{align}

Construction of the abnormal returns for the generalised synthetic control involves taking the difference between observed returns and those of the counterfactual version of that share. Consequently comparison can only be done on those shares considered ``treated'' by listing to, or being de-listed from, the DJSI. In the subsequent sections we report the CAPM CARs at an aggregate level and broken down by industry-year for those joining firms. Note further that because those announced as either gaining listing on, or being de-listed from, the DJSI are included in both the full and base samples there is no distinction between samples in the later reporting.

\subsection{Descriptive Statistics}
\label{sec:desc}

\begin{table}
	\begin{center}
	\begin{tiny}
		\caption{Descriptive Statistics \label{tab:sumstat}}
		\begin{tabular}{c l c c c c c c c c}
			\multicolumn{10}{l}{Panel A: Summary Statistics for Full Sample (Listing)}\\
			&& Mean & Min & 25th pctile & Median & 75th pctile & Max & St. dev. & $N$\\
			\hline
			(1)&DJSI & 0.008 & 0.000 & 0.000 & 0.000 & 0.000 & 1 & 0.089 & 23952 \\
			(2)&Size & 7.392 & 1.548 & 6.053 & 7.284 & 8.535 & 14.76 & 1.850 & 23592 \\
			(3)&Profitability & 0.088 & -11.71 & 0.045 & 0.101 & 0.167 & 13.65 & 7.003 & 22458\\
			(4)&Leverage & 0.432 & -24.78 & 0.150 & 0.391 & 0.651 & 11.55 & 0.536 & 23291 \\
			(5)&CAR[1,16]& 0.115 & -89.18 & -4.146 & -0.324 & 3.743 & 128.1 & 8.795 & 23592\\
			(6)&CAR[1,21]& 0.144 & -81.00 & -4.735 & -0.261 & 4.422 & 120.0 & 9.963 & 23592\\
			(7)&CAR[16,21]& 0.083 & -94.76 & -2.252 & -0.025 & 2.331 & 126.2 & 5.182 & 23592\\
			(8)&CAR[16,36]& -0.047 & -112.1 & -4.823 & 0.060 & 4.906 & 151.7 & 10.42 & 23592\\
			(9)&CAR[21,36]& -0.180 & -97.76 & -4.218 & -0.027 & 4.053 & 165.4 & 9.073 & 23592 \\
			\hline
			\multicolumn{10}{l}{Panel B: Summary Statistics for Base Sample (Listing)}\\
			\hline
			(10) & DJSI & 0.047 & 0.000 & 0.000 & 0.000 & 0.000 & 1 & 0.197 & 4020 \\
			(11) & Size & 9.728 & 6.617 & 8.678 & 9.556 & 10.54 & 14.76 & 1.552 & 4020 \\
			(12) & Profitability & 0.142 & -2.115 & 0.068 & 0.123 & 0.196 & 5.279 & 0.228 &3888 \\
			(13) & Leverage & -0.180 & -13.38 & 0.352 & 0.504 & 0.724 & 8.369 & 0.405 & 3949 \\
			
			\hline
			\multicolumn{10}{l}{Panel C: Univariate sample comparisons (Listing)}\\
			\hline
			&& \multicolumn{3}{l}{Full Sample} & \multicolumn{2}{l}{} & \multicolumn{3}{l}{Base Sample}\\
			&& List & Other & Diff. & & & List & Other & Diff.\\
			\hline
			
			\hline(14) & Size & 10.02 & 7.370 & 2.650*** & & Size & 10.02 & 9.713 & 0.307**\\
			(15) & Profitability & 0.196 & 0.087 & 0.109*** & &Profitability & 0.196 & 0.139 & 0.057\\
			(16) & Leverage & 0.524 & 0.431 & 0.093*** & & Leverage & 0.524 & 0.434 &0.090 \\
			\multicolumn{10}{l}{Panel D: Correlations (Listing)}\\
			\hline
			&& \multicolumn{4}{l}{Full Sample}  & \multicolumn{4}{l}{Base Sample}\\
			&& DJSI & Size & Profit & Leverage & DJSI & Size & Profit & Leverage\\
			(17)&DJSI & 1 & & & & 1 & & &\\
			(18)&Size & 0.131 & 1 & & & 0.045 & 1 & & \\
			(19)&Profit & 0.031 & 0.149 & 1 & & 0.060 & 0.042 & 1 & \\
			(20)&Leverage & 0.027 & 0.504 & 0.056 & 1 & -0.013 & 0.274 & 0.041 & 1 \\
			\hline
			\multicolumn{10}{l}{Panel E: Summary Statistics for Full Sample (De-Listing)}\\
			&& Mean & Min & 25th pctile & Median & 75th pctile & Max & St. dev. & $N$\\
			\hline
			(21)&DJSIX & 0.006 & 0.000 & 0.000 & 0.000 & 0.000 & 1 & 0.078 & 16746 \\
			(22)&Size & 7.502 & 1.748 & 6.160 & 7.400 & 8.649 & 14.76 & 1.854 & 16746 \\
			(23)&Profitability & 0.073 & -21.62 & 0.036 & 0.093 & 0.159 & 7.003 & 0.371 & 16746\\
			(24)&Leverage & 0.432 & 0.000 & 0.172 & 0.411 & 0.654 & 0.999 & 0.293 & 16746 \\
			(25)&CAR[1,16]& 0.336 & -68.41 & -4.110 & -0.161 & 3.993 & 140.9 & 8.943 & 16746\\
			(26)&CAR[1,21]& 0.269 & -66.69 & -4.621 & -0.137 & 4.560 & 124.0 & 10.00 & 16746\\
			(27)&CAR[16,21]& -0.095 & -68.43 & -2.490 & -0.067 & 2.341 & 111.6 & 5.232 & 16746\\
			(28)&CAR[16,36]& -0.190 & -112.7 & -4.816 & 0.139 & 4.867 & 152.7 & 10.59& 16746\\
			(29)&CAR[21,36]& -0.219 & -98.53 & -4.074 & 0.209 & 4.167 & 167.6 & 9.411 & 16746 \\
			\hline
			\multicolumn{10}{l}{Panel F: Summary Statistics for Base Sample (De-Listing)}\\
			\hline
			(30) & DJSIX & 0.040 & 0.000 & 0.000 & 0.000 & 0.000 & 1 & 0.195 & 2612 \\
			(31) & Size & 10.05 & 7.245 & 9.057 & 9.924 & 10.82 & 14.76 & 1.367 & 2612 \\
			(32) & Profitability & 0.135 & -1.235 & 0.062 & 0.114 & 0.193 & 5.279 & 0.239 & 2612 \\
			(33) & Leverage & 0.536 & 0.006 & 0.363 & 0.517 & 0.726 & 0.995 & 0.079 & 2612 \\
			
			\hline
			\multicolumn{10}{l}{Panel G: Univariate sample comparisons (De-Listing)}\\
			\hline
			&& \multicolumn{3}{l}{Full Sample} & \multicolumn{2}{l}{} & \multicolumn{3}{l}{Base Sample}\\
			&& List & Other & Diff. & & & List & Other & Diff.\\
			\hline
			
			\hline(34) & Size & 10.83 & 7.500 & 3.329*** & & Size & 10.83 & 10.05 & 0.784*\\
			(35) & Profitability & 0.157 & 0.073 & 0.084 & &Profitability & 0.157 & 0.153 & 0.022\\
			(36) & Leverage & 0.565 & 0.432 & 0.133* & & Leverage & 0.565 & 0.536 &0.029 \\
			\multicolumn{10}{l}{Panel H: Correlations (De-Listing)}\\
			\hline
			&& \multicolumn{4}{l}{Full Sample}  & \multicolumn{4}{l}{Base Sample}\\
			&& DJSIX & Size & Profit & Leverage & DJSIX & Size & Profit & Leverage\\
			(37)&DJSIX & 1 & & & & 1 & & &\\
			(38)&Size & 0.119 & 1 & & & 0.037 & 1 & & \\
			(39)&Profit & 0.025 & 0.146 & 1 & & 0.047 & 0.005 & 1 & \\
			(40)&Leverage & 0.026 & 0.503 & 0.045 & 1 & -0.006 & 0.235 & 0.009 & 1 \\
			\hline
		\end{tabular}
	\end{tiny}
	\end{center}
\raggedright
\footnotesize{Notes: Descriptive statistics for variables used in main analyses. Samples are restricted by two digit NAICS code to those industries with one or more firm joining the DJSI within a given year. Full sample includes all firms listed on the major American stock exchanges with sufficient data, with the base sample considering only those with assets at least 80\% as large as the smallest joining firm in their industry. All stock data is sourced from CRSP. DJSI listing data is taken from Robecco SAM. Size (log assets), profitability(return on equity) and leverage (ratio of total debt to total capital) are sourced from Compustat. Significance given by * = 10\%, ** = 5\%, *** = 1\%.}
\end{table}

Descriptive statistics in Table \ref{tab:sumstat} remind just how few firms obtain listing within any given year; just 0.8\% of all observations represent listings. Focusing only on larger firms in Panel B that figure rises to 4.7\%, but this is still low relative to the overall volume of data. A full breakdown of the entrants by NAICS2 code is provided in Table \ref{tab:tandc}. Rows (2) to (4) provide some statistics for three key firm characteristics. Size, measured as the log of total assets, shows that firms within the full sample are drawn from a wide ranging distribution. This diversity is the motivation for the reduction in the sample. The base sample has a much higher average and a minimum value close to the median of the full sample. Profitability, captured as the return on equity, in rows (3) and (12), is also wide ranging with a number of firms reporting losses in both samples. Once the minimum size requirement is imposed the minimum ROE is much larger. Leverage also has a smaller range amongst the largest firms. Comparison of means on rows (4) and (13) verify this pattern. 

Focus in this paper is on the abnormal returns, if any, gained when entering the DJSI. For this purpose CAR are used, calculated using \eqref{eq:car2}. For the full sample, rows (5) to (9) give values for five periods of interest. From the start of the treatment period to announcement date, days 1 to 16, we note a small positive abnormal return of 0.115\% amongst the whole sample. From the first day to the effective day this average has increased to 0.144\%. Within the week from announcement date to the effective date there are thus positive abnormal returns to be had. Row (7) shows these to be 0.083\%. Looking at periods beginning on the two key listing dates, announcement on day 16 and effective on day 21, to the end of the sample the CARs are -0.047\% and -0.180\% on average. From the positive pre-announcement and negative post effective date in particular we see much of the pre-announcement and correction effects discussed within the literature. As these figures contain all firms they remind that there will be many more stories behind the results, and that it is not possible to attribute all of these changes to the DJSI listings. 

Univariate tests in Panel C inform on the differences between those firms who join the DJSI and the control groups for that given year. These are aggregated into a large list and tested for equality of mean between the joining and non-joining samples. In both the full sample and the base sample the firms joining the list are significantly larger, this result remains even when the restriction based on size has been imposed. Looking at profitability the joining firms have a significantly higher ROE than the non-joiners; this would be consistent with the broad observation that improving CSR is costly and therefore typically only practised by firms who have the profitability to support such measures. After reducing the sample to the base, the average ROE for control firms rises but the gap between treatment and control is no longer significant. Finally, we see that amongst the whole sample firms joining the DJSI have a higher debt to capital ratio, but in the base sample it is the non-joining firms who have the higher leverage. The latter difference is also not significant however. It is therefore suggested that the largest firms with the greater profitability and ability to raise their leverage to fund investment in projects which will raise sustainability performance, who are most likely to join. 

Panel D addresses the correlations between the data. Leverage and size are the most correlated, but fall short of the 0.7 threshold usually assumed to be a concern for multicollinearity. For the base sample the correlations between the three financial variables drop significantly. Correlations between DJSI listing and all three controls are low in both the full, and base, samples. Thus in any regression contexts where these variables feature we can have confidence in the inference gained. 

Relative to the listing sample the de-listing sample is much smaller, there are just 16746 firm-years instead of the previous 23592. A de-listing proportion of just 0.6\% is lower than the 0.8\% joining proportion. Since its' inception in 2005 the DJSI North America has been growing in numbers; a larger proportion of joining firms is therefore to be expected. Summary statistics for firm characteristics in rows (22) to (24) are very similar to those in Panel A, the de-listing sample being slightly larger, slightly less profitable but being of identical leverage on average. Comparing the CARs for the periods summarised in lines (25) to (29), shows that those firms which are to leave the market return an average of 0.336\% prior to the announcement compared to just 0.115\% for those destined to gain listing. Negative effects post listing are also larger in absolute value for the de-listing group. Biggest contrast comes in the period between announcement day and the day the changes become effective. Returns between announcement and effective dates in the listing case were 0.083\% whilst the de-listing set showed losses at 0.095\%. Because there is great variation across years and firms it is not instructive to read deeply into this, but there is a suggestion of listing and de-listing moving in the opposite directions as intuition would suggest. Reducing the sample to consider only those control firms with assets of at least 80\% of those of the smallest firm that gets de-listed in their industry-year produces a set of just 2612 observations. Summary statistics for this base sample, lines (30) to (33), have similar properties to those for the listing case (lines 10-13). 

T-tests of firm characteristics between de-listed firms and their respective controls indicate that those leaving the DJSI were larger and more highly leveraged than others. Unlike the listing case there is no significance to the profitability differential. Combined with the lower average, 0.157 for de-listing firms versus 0.196 for those joining, there is indication towards the greater profitability of being seen as a CSR leader through the recognition afforded by remaining a DJSI member. Such is only indicative since it relies on consumers knowing in the previous financial year that the de-listed firms were not performing as strongly as those who were to gain listing\footnote{Consider the chronological ordering required. CSR reputation is observed by consumers who then make purchasing decisions accordingly. These purchasing decisions affect sales, and hence profits. To be recognised within the data here such changes would have to be seen in the financial variables more than nine months before the announcement is made. Such is not unreasonable since in many industries it is possible to know who the likely listees will be, or who the de-listed firms are likely to be, well ahead. Such a chronology may thus not be universal and so the temptation to generalise the motivation for the varied profitability is left as an intuition.}. Panel G further shows that even after reduction to the Base Sample size differentials are still significant. Correlation statistics in Panel H again urge caution on the relationship between size and leverage. In all tests performed on the regressions this is not seen as a problem to the reliability of the results that follow; maximum correlation is again shy of the 0.7 that would be problematic, for example.

\section{DJSI Membership and Stock Returns}
\label{sec:first}

Identification of membership effects to date has relied on the comparison of samples or a dummy variable regression approach. Our preliminary discussion thus considers these two approaches, presenting univariate tests of equality in returns and CAR for periods surrounding the DJSI listing announcement date. Event studies proceed in this way, but may opt to perform matching between samples prior to conducting the univariate tests. For this reference case we maintain the maximal set of data and do not exclude firms for which sufficient data is available. Second we consider common company financial variables that have been linked to stock returns. Thus, in the spirit of \cite{acemoglu2016value} we identify a measure of the listing, and de-listing, effects which controls for these typically considered determinants. Both approaches corroborate the inconclusiveness identified within the literature. 

\subsection{Two-Sample Approach}
\label{sec:twosamp}

Identifying listing effects by comparing samples of firms entering the DJSI with their peers reveals many of the already identified phenomenon. First we consider the posted excess returns for the share sample. Whilst not accounting for past performance these do deliver the most direct outward impression of the performance of the shares of the new entrants to the DJSI. Compared to larger firms new entrants deliver significantly lower returns on day 9, 5 trading days prior to the announcement. On the announcement day listed firms offer a return 0.3\% higher than non-listed and 0.2\% higher than the largest control firms. Only a lower abnomral return on days 20 and 36 for newly listed firms compared to all others is significant. Such a lack of impact is suggestive that the CAPM model is pricing the DJSI listed firms with reasonable accuracy. Event studies have spoken of the positive pre-announcement effect and for those firms who are to gain listing we do see evidence of such here, although many of the positive differences are not significant.

\begin{table}
	\begin{center}
		\caption{Univariate Tests of Return Equality \label{tab:uni}}
			\begin{tiny}
			\begin{tabular}{l c c c c c c c c c c}
				\hline
				
				Day & \multicolumn{5}{l}{Listing} & \multicolumn{5}{l}{Delisting} \\ 
				& $DJSI^{+}$ & \multicolumn{2}{l}{All firms} & \multicolumn{2}{l}{Base Sample} & $DJSI^{-}$ & \multicolumn{2}{l}{All firms} & \multicolumn{2}{l}{Base Sample} \\
				& Return & Return & Diff & Return & Diff & Return & Return & Diff & Return & Diff \\
				\hline
				1&-0.010&0.145&-0.155&0.031&-0.041& -0.129 & 0.188 & -0.316*** & 0.011 & -0.139\\
				2&0.086&-0.03&0.115&-0.019&0.105 & -0.214 & -0.012 & -0.202* & -0.068 & -0.146\\
				3&0.012&0.077&-0.065&-0.071&0.083 & -0.160 & 0.095 & -0.256** & -0.014 & -0.146\\
				4&-0.049&-0.113&0.064&-0.074&0.025 & 0.039 & -0.201 & 0.239** & -0.217 & 0.255**\\
				5&-0.078&-0.016&-0.062&-0.019&-0.058 & -0.085 & -0.035 & -0.049 & 0.036 & -0.120\\
				6&0.049&0.061&-0.012&-0.009&0.058 & -0.073 & 0.079 & -0.152 & 0.011 & -0.084\\
				7&-0.11&0.022&-0.132&-0.087&-0.023 & 0.189 & 0.091 & 0.098 & -0.154 & 0.342***\\
				8&-0.011&-0.002&-0.008&-0.013&0.002 & 0.044 & 0.073 & -0.029 & 0.047 & -0.003\\
				9&0.129&-0.111&0.239*&0.011&0.118 & 0.230 & -0.161 & 0.392*** & -0.096 & 0.327**\\
				10&0.066&-0.034&0.099&0.029&0.036 & 0.180 & -0.001 & 0.182 & 0.044 & 0.136\\
				11&0.095&0.114&-0.019&-0.001&0.096&0.071 & 0.170 & -0.099 & -0.044 & 0.115\\
				12&-0.195&-0.05&-0.145&-0.103&-0.092& 0.017 & -0.028 & 0.044 & -0.135 & 0.152\\
				13&-0.04&0.057&-0.096&0.073&-0.113 & 0.048 & 0.183 & -0.135 & -0.036 & 0.084 \\
				14&-0.198&-0.095&-0.104&-0.114&-0.085 & -0.120 & -0.098 & -0.022 & -0.182 & 0.061 \\
				15&0.077&0.07&0.007&0.076&0.001 & -0.121 & 0.024 & -0.145 & -0.139 & 0.018\\
				ANN&0.236&-0.064&0.299***&0.01&0.226** & 0.033 & -0.147 & 0.180 & -0.104 & 0.137\\
				17&-0.046&-0.133&0.087&-0.145&0.099 & 0.226 & -0.068 & 0.184 & 0.106 & 0.010\\
				18&0.086&-0.003&0.09&0.091&-0.004 & -0.043 & -0.133 & 0.090 & -0.016 & -0.026\\
				19&0.095&-0.064&0.159&-0.035&0.131 & -0.091 & -0.050 & -0.041 & -0.108 & 0.016\\
				20&0.010&0.268&-0.258*&0.050&-0.041 & -0.006 & 0.426 & -0.432** & 0.053 & -0.058\\
				21&-0.075&-0.086&0.01&-0.055&-0.020 & 0.017 & -0.125 & 0.142 & -0.002 & 0.019\\
				22&-0.174&-0.071&-0.103&0.008&-0.182 & 0.054 & -0.109 & 0.163 & 0.016 & 0.039\\
				23&0.001&0.092&-0.090&-0.146&0.147 & -0.135 & 0.155 & -0.290 & -0.044 & -0.092\\
				24&-0.031&-0.088&0.057&-0.114&0.083 & 0.111 & 0.008 & 0.103 & -0.057 & 0.168\\
				25&-0.089&0.009&-0.098&-0.050&-0.039 & -0.108 & -0.041 & -0.066 & -0.005 & -0.103\\
				26&-0.076&-0.027&-0.049&-0.019&-0.057 & 0.139 & 0.062 & 0.077 & -0.002 & 0.140\\
				27&-0.033&-0.099&0.066&0.030&-0.063 & 0.012 & -0.056 & 0.068 & 0.018 & -0.006\\
				28&-0.162&0.036&-0.198&0.006&-0.168 & 0.354 & 0.034 & 0.320* & 0.020 & 0.334**\\
				29&0.035&0.168&-0.133&0.029&0.006 & -0.006 & 0.260 & -0.266* & -0.109 & 0.103\\
				30&-0.059&-0.169&0.111&-0.089&0.031 & -0.303 & -0.173 & -0.131 & 0.079 & -0.382\\
				31&-0.171&0.014&-0.186&-0.051&-0.120 & -0.149 & 0.047 & -0.196 & 0.042 & -0.191\\
				32&0.003&-0.037&0.039&-0.081&0.084 & 0.001 & 0.012 & -0.011 & 0.009 & -0.008\\
				33&-0.021&0.071&-0.092&-0.122&0.101 & -0.045 & 0.188 & -0.233 & 0.024 & -0.069\\
				34&0.123&-0.123&0.246&-0.084&0.207&0.432 & -0.221 & 0.653*** & 0.029 & 0.403**\\
				35&0.021&-0.131&0.152&-0.164&0.185&0.008 & -0.177 & 0.185 & -0.079 & 0.087\\
				36&-0.102&0.100&-0.202*&-0.021&-0.081 & -0.202 & -0.084 & -0.118 & -0.278 & 0.076\\
				\hline
			\end{tabular}
		\end{tiny}
	\end{center}
\raggedright
\footnotesize{Notes: Abnormal returns are calculated based on the difference between realised excess returns and the fitted value using coefficients estimated individually for each firm during the control period. $DJSI^{+}$ refers to firms which join the DJSI, $DJSI^{-}$ being those who are de-listed. All firms include any share listed on the major US exchanges from the same industry as a joining firm, with large firms including only those with assets 80\% of those of the smallest new entrant to/exiting firm from the DJSI. Evaluation processes are repeated annually such that reported figures represent the average effect across the period. In the returns case period represents the trading day for which the returns are reported. Diff reports the difference between the treated firms, listed or de-listed, and the untreated firms in the appropriate sample. Asterisks denote significance levels of a two-tailed t-test (*** - 1\%, ** - 5\% and * - 1\%).}
\end{table}

In the delisting effect there is more significance particular further away from the key announcement and effective dates. Attributing this to the DJSI is harder, but this is a difference between de-listed firms and others. That the effect is of a similar magnitude when comparing with large firms as it is when comparing with the full sample means it is not a size related effect. Moving into the period a week before the announcement there are no significant differentials between those who will be dropped from the index and any of the control groups. A positive effect here can be aligned to the traditionalist perspective that CSR is an expensive luxury for firms that is better reduced. Higher profitability would be expected from de-listing and therefore this uptick is consistent with expectation of better future performance post delisting. There is some evidence of a correction effect moving against these positive returns; values are similar in the delisting columns as they are in the listing case. Contrasts between the two change directions are stark but should not be viewed as demonstrative of a lack of appreciation of the role of DJSI membership. Heterogeneities in investor attitude are manifesting through revisionists driving returns on listed firms up and others who favour the traditionalist perspective being behind the boost to de-listing firms. Coexistence of the two effects is in line with \cite{oberndorfer2013does}.

\begin{table}[htbp]
	\begin{center}
		\begin{tiny}
	\caption{Cumulative Abnormal Returns t-test Summaries: Listing}
	\begin{tabular}{llllllllllll}
		\hline
		\multicolumn{12}{l}{Panel (a): Full Sample ($N=22458$):}\\
		From/To  & 11    & 12    & 13    & 14    & 15    & ANN & 17    & 18    & 19    & 20    & EFF \\
		1     & 0.065 & -0.08 & -0.176 & -0.28 & -0.273 & 0.026 & 0.113 & 0.202 & 0.362 & 0.104 & 0.114 \\
		2     & 0.22  & 0.075 & -0.021 & -0.125 & -0.118 & 0.181 & 0.268 & 0.358 & 0.517 & 0.259 & 0.269 \\
		3     & 0.104 & -0.04 & -0.137 & -0.24 & -0.234 & 0.066 & 0.152 & 0.242 & 0.402 & 0.144 & 0.154 \\
		4     & 0.17  & 0.025 & -0.071 & -0.175 & -0.168 & 0.131 & 0.218 & 0.308 & 0.467 & 0.209 & 0.219 \\
		5     & 0.106 & -0.039 & -0.135 & -0.239 & -0.232 & 0.067 & 0.154 & 0.244 & 0.403 & 0.145 & 0.155 \\
		6     & 0.168 & 0.023 & -0.073 & -0.177 & -0.17 & 0.129 & 0.216 & 0.306 & 0.465 & 0.207 & 0.217 \\
		7     & 0.179 & 0.035 & -0.062 & -0.165 & -0.159 & 0.141 & 0.228 & 0.317 & 0.477 & 0.219 & 0.229 \\
		8     & 0.312 & 0.167 & 0.071 & -0.033 & -0.026 & 0.273 & 0.36  & 0.45  & \multicolumn{1}{l}{0.609*} & 0.351 & 0.361 \\
		9     & \multicolumn{1}{l}{0.32*} & 0.175 & 0.079 & -0.025 & -0.018 & 0.281 & 0.368 & 0.458 & \multicolumn{1}{l}{0.617*} & 0.359 & 0.37 \\
		10    & 0.081 & -0.064 & -0.16 & -0.264 & -0.257 & 0.042 & 0.129 & 0.219 & 0.378 & 0.12  & 0.131 \\
		11    & -0.019 & -0.163 & -0.26 & -0.363 & -0.357 & -0.057 & 0.03  & 0.119 & 0.279 & 0.021 & 0.031 \\
		12    &       & -0.145 & -0.241 & \multicolumn{1}{l}{-0.345*} & -0.338 & -0.038 & 0.048 & 0.138 & 0.297 & 0.039 & 0.05 \\
		13    &       &       & -0.096 & -0.2  & -0.193 & 0.106 & 0.193 & 0.283 & 0.442 & 0.184 & 0.194 \\
		14    &       &       &       & -0.104 & -0.097 & 0.203 & 0.289 & \multicolumn{1}{l}{0.379*} & \multicolumn{1}{l}{0.538*} & 0.28  & 0.291 \\
		15    &       &       &       &       & 0.007 & \multicolumn{1}{l}{0.306*} & \multicolumn{1}{l}{0.393**} & \multicolumn{1}{l}{0.483**} & \multicolumn{1}{l}{0.642***} & 0.384 & 0.394 \\
		ANN &       &       &       &       &       & \multicolumn{1}{l}{0.299***} & \multicolumn{1}{l}{0.386***} & \multicolumn{1}{l}{0.476***} & \multicolumn{1}{l}{0.635***} & 0.377 & 0.388 \\
		17    &       &       &       &       &       &       & 0.087 & 0.176 & \multicolumn{1}{l}{0.336*} & 0.078 & 0.088 \\
		18    &       &       &       &       &       &       &       & 0.09  & 0.249 & -0.009 & 0.002 \\
		19    &       &       &       &       &       &       &       &       & 0.159 & -0.099 & -0.088 \\
		20    &       &       &       &       &       &       &       &       &       & \multicolumn{1}{l}{-0.258*} & \multicolumn{1}{l}{-0.247*} \\
		EFF &       &       &       &       &       &       &       &       &       &       & 0.01 \\
		&\\
		From/To & EFF & 22     & 24      & 26    & 31    & 36    &      && &       &  \\
		11    & 0.031 & -0.072  & -0.105  & -0.253 & -0.593 & -0.449 & &&      &       &  \\
		12    & 0.05  & -0.053  & -0.087 & -0.235 & -0.574 & -0.43 &    &&   &       &  \\
		13    & 0.194 & 0.091 &0.058& -0.090 & -0.429 & -0.286 &       &  &&     &  \\
		14    & 0.291 & 0.188  & 0.154  & 0.006 & -0.333 & -0.189 &       &    &&   &  \\
		15    & 0.394 & 0.291  & 0.258 & 0.110  & -0.229 & -0.086 &       &      && &  \\
		ANN & 0.388 & 0.284  & 0.251  & 0.103 & -0.236 & -0.092 &       &       &  &&\\
		17    & 0.088 & -0.015 & -0.048 & -0.196 & -0.536 & -0.392 &       &     &&  &  \\
		18    & 0.002 & -0.102  & -0.135 & -0.283 & -0.622 & -0.479 &       &   &&    &  \\
		19    & -0.088 & -0.191 & -0.225 & -0.372 & -0.712 & -0.568 &       &    &&   &  \\
		20    & \multicolumn{1}{l}{-0.247*} & \multicolumn{1}{l}{-0.351*}  & -0.384 & -0.532 & \multicolumn{1}{l}{-0.871*} & -0.728 &       &       & && \\
		EFF & 0.010  & -0.093  & -0.126 & -0.274 & -0.613 & -0.470 &    &&   &       &  \\
		22    &       & -0.103  & -0.136  & -0.284 & -0.624 & -0.480 &&&       &       &  \\
		\hline
		\multicolumn{12}{l}{Panel (b): Base Sample ($N=3888$):}\\
		From /To & 11    & 12    & 13    & 14    & 15    & ANN & 17    & 18    & 19    & 20    & EFF \\
		1     & 0.4   & 0.308 & 0.196 & 0.111 & 0.112 & 0.337 & 0.437 & 0.432 & 0.563 & 0.522 & 0.502 \\
		2     & 0.441 & 0.349 & 0.237 & 0.152 & 0.153 & 0.378 & 0.478 & 0.473 & 0.604 & 0.563 & 0.543 \\
		3     & 0.336 & 0.244 & 0.132 & 0.047 & 0.048 & 0.273 & 0.373 & 0.368 & 0.499 & 0.458 & 0.438 \\
		4     & 0.253 & 0.162 & 0.049 & -0.036 & -0.035 & 0.191 & 0.29  & 0.286 & 0.416 & 0.376 & 0.356 \\
		5     & 0.229 & 0.137 & 0.024 & -0.06 & -0.06 & 0.166 & 0.265 & 0.261 & 0.392 & 0.351 & 0.331 \\
		6     & 0.287 & 0.195 & 0.083 & -0.002 & -0.001 & 0.224 & 0.324 & 0.319 & 0.45  & 0.409 & 0.389 \\
		7     & 0.229 & 0.138 & 0.025 & -0.06 & -0.059 & 0.167 & 0.266 & 0.262 & 0.392 & 0.352 & 0.331 \\
		8     & 0.252 & 0.161 & 0.048 & -0.037 & -0.036 & 0.19  & 0.289 & 0.285 & 0.415 & 0.375 & 0.354 \\
		9     & 0.25  & 0.158 & 0.046 & -0.039 & -0.038 & 0.187 & 0.287 & 0.282 & 0.413 & 0.372 & 0.352 \\
		10    & 0.132 & 0.041 & -0.072 & -0.157 & -0.156 & 0.07  & 0.169 & 0.165 & 0.295 & 0.255 & 0.234 \\
		11    & 0.096 & 0.004 & -0.109 & -0.193 & -0.193 & 0.033 & 0.132 & 0.128 & 0.259 & 0.218 & 0.198 \\
		12    &       & -0.092 & -0.205 & -0.289 & -0.289 & -0.063 & 0.036 & 0.032 & 0.163 & 0.122 & 0.102 \\
		13    &       &       & -0.113 & -0.197 & -0.197 & 0.029 & 0.128 & 0.124 & 0.255 & 0.214 & 0.194 \\
		14    &       &       &       & -0.085 & -0.084 & 0.142 & 0.241 & 0.237 & 0.368 & 0.327 & 0.307 \\
		15    &       &       &       &       & 0.001 & 0.226 & \multicolumn{1}{l}{0.326*} & \multicolumn{1}{l}{0.321*} & \multicolumn{1}{l}{0.452*} & 0.411 & 0.391 \\
		\multicolumn{1}{l}{ANN} &       &       &       &       &       & \multicolumn{1}{l}{0.226**} & \multicolumn{1}{l}{0.325**} & \multicolumn{1}{l}{0.321*} & \multicolumn{1}{l}{0.452*} & 0.411 & 0.391 \\
		17    &       &       &       &       &       &       & 0.099 & 0.095 & 0.226 & 0.185 & 0.165 \\
		18    &       &       &       &       &       &       &       & -0.004 & 0.127 & 0.086 & 0.066 \\
		19    &       &       &       &       &       &       &       &       & 0.131 & 0.09  & 0.07 \\
		20    &       &       &       &       &       &       &       &       &       & -0.041 & -0.061 \\
		\multicolumn{1}{l}{EFF} &       &       &       &       &       &       &       &       &       &       & -0.02 \\
		&\\
		& \multicolumn{1}{l}{EFF} & 22  & 24    & 26    & 31    & 36    &  &&     &       &  \\
		11    & 0.198 & 0.016  & 0.246 & 0.150  & -0.164 & 0.331 &   &&    &       &  \\
		12    & 0.102 & -0.080 & 0.150  & 0.054 & -0.26 & 0.236 &  &&     &       &  \\
		13    & 0.194 & 0.012 & 0.242 & 0.146 & -0.168 & 0.327 &     &&  &       &  \\
		14    & 0.307 & 0.125  & 0.355 & 0.259 & -0.055 & 0.44  &  &&     &       &  \\
		15    & 0.391 & 0.210 & 0.440   & 0.344 & 0.03  & 0.525 &     &&  &       &  \\
		\multicolumn{1}{l}{ANN} & 0.391 & 0.209 & 0.439 & 0.343 & 0.029 & 0.524 & &&      &       &  \\
		17    & 0.165 & -0.017  & 0.213 & 0.117 & -0.197 & 0.298 & &&      &       &  \\
		18    & 0.066 & -0.116  & 0.114   & 0.018 & -0.296 & 0.199 &   &&    &       &  \\
		19    & 0.07  & -0.112  & 0.118  & 0.022 & -0.292 & 0.203 &     &&  &       &  \\
		20    & -0.061 & -0.243  & -0.013 & -0.109 & -0.423 & 0.073 &   &&    &       &  \\
		\multicolumn{1}{l}{EFF} & -0.020 & -0.202 & 0.028 & -0.068 & -0.382 & 0.113 & &&      &       &  \\
		22    &       & -0.182 & 0.048 & -0.048 & -0.362 & 0.133 &       &  &&     &  \\
		
		\hline
	\end{tabular}%
	\label{tab:carsp}%
	\end{tiny}
	\end{center}
\raggedright
\footnotesize{Notes: Cumulative abnormal returns are calculated using the fitted values from the CAPM. Full Sample includes all firms that did not join the DJSI in any industry-year where one or more firms did join. Base Sample restricts the full sample to include all joining firms and only those non-joining firms with an asset holding at least 80\% of that of the lowest entrant in their industry-year. Significance given for a two-sample t-test of equality between joining and non-joining firms, with * - 5\%, ** - 1\%, *** - 0.1\%}
\end{table}%

\begin{table}[htbp]
	\begin{center}
		\begin{tiny}
			\caption{Cumulative Abnormal Returns t-test Summaries: De-Listing}
			\begin{tabular}{llllllllllll}
				\hline
				\multicolumn{12}{l}{Panel (a): Full Sample ($N=16747$):}\\
				From/To   & 11 & 12 & 13 & 14 & 15 & ANN & 17 & 18 & 19 & 20 & EFF\\
				1 &   -0.193&-0.148&-0.284&-0.305&-0.45&-0.271&-0.087&0.004&-0.038&-0.47&-0.328\\
				2&0.124&0.168&0.033&0.011&-0.134&0.046&0.230&0.320&0.279&-0.153&-0.011\\
				3&0.325&0.37&0.234&0.213&0.068&0.247&0.431&0.522&0.48&0.048&0.190\\
				4&0.581&0.625&0.49&0.468&0.323&0.503&0.687&0.777&0.736&0.304&0.446\\
				5&0.342&0.386&0.251&0.229&0.084&0.264&0.448&0.538&0.497&0.065&0.207\\
				6&0.391&0.435&0.3&0.278&0.133&0.313&0.497&0.587&0.546&0.114&0.256\\
				7&0.543*&0.587*&0.452&0.43&0.285&0.465&0.649&0.739*&0.698*&0.266&0.408\\
				8&0.445*&0.489&0.354&0.332&0.187&0.367&0.551&0.641&0.6&0.168&0.310\\
				9&0.474**&0.518*&0.383&0.361&0.216&0.396&0.58&0.67&0.629&0.197&0.339\\
				10&0.082&0.127&-0.008&-0.030&-0.175&0.004&0.188&0.279&0.237&-0.195&-0.053\\
				11&-0.099&-0.055&-0.190&-0.212&-0.357&-0.177&0.007&0.097&0.056&-0.376&-0.234\\
				12&&0.044&-0.091&-0.113&-0.258&-0.078&0.106&0.196&0.155&-0.277&-0.135\\
				13&&&-0.135&-0.157&-0.302&-0.122&0.062&0.152&0.111&-0.321&-0.179\\
				14&&&&-0.022&-0.167&0.013&0.197&0.287&0.246&-0.186&-0.044\\
				15&&&&&-0.145&0.035&0.219&0.309&0.268&-0.164&-0.022\\
				ANN&&&&&&0.180&0.364**&0.454**&0.413&-0.019&0.122\\
				17&&&&&&&0.184&0.274&0.233&-0.199&-0.057\\
				18&&&&&&&&0.09&0.049&-0.383&-0.241\\
				19&&&&&&&&&-0.041&-0.473**&-0.331\\
				20&&&&&&&&&&-0.432***&-0.290\\
				EFF&&&&&&&&&&&0.142\\
				From/To & EFF & 22 & 24 & 26 & 31 & 36&&&&&\\
				11&-0.234&-0.071&-0.361&-0.248&-0.453&0.024&&&&&\\
				12&-0.135&0.028&-0.262&-0.149&-0.353&0.123&&&&&\\
				13&-0.179&-0.017&-0.307&-0.193&-0.398&0.079&&&&&\\
				14&-0.044&0.119&-0.171&-0.058&-0.263&0.214&&&&&\\
				15&-0.022&0.14&-0.15&-0.036&-0.241&0.236&&&&&\\
				ANN&0.122&0.285&-0.005&0.109&-0.096&0.381&&&&&\\
				17&-0.057&0.106&-0.184&-0.071&-0.276&0.201&&&&&\\
				18&-0.241&-0.078&-0.368&-0.255&-0.459&0.017&&&&&\\
				19&-0.331&-0.169&-0.459&-0.345&-0.55&-0.073&&&&&\\
				20&-0.290&-0.127&-0.417&-0.304&-0.508&-0.032&&&&&\\
				EFF&0.142&0.305&0.015&0.128&-0.076&0.400&&&&&\\
				22&&0.163&-0.127&-0.014&-0.218&0.258&&&&&\\
				\multicolumn{12}{l}{Panel (b): Base Sample ($N=2612$):}\\
				From/To  & 11 & 12 & 13 & 14 & 15 & ANN & 17 & 18 & 19 & 20 & EFF \\
				1&0.538&0.690&0.774&0.835&0.853&0.991*&1.000*&0.974*&0.990*&0.932&0.951*\\
				2&0.677&0.830*&0.913*&0.975**&0.993**&1.130**&1.140**&1.113**&1.130**&1.071*&1.090**\\
				3&0.823**&0.975**&1.059**&1.120**&1.138**&1.276***&1.285**&1.259**&1.276**&1.217**&1.236**\\
				4&0.969***&1.121***&1.205***&1.266***&1.285***&1.422***&1.432***&1.405***&1.422***&1.363**&1.382***\\
				5&0.714**&0.866**&0.95**&1.011**&1.029**&1.166***&1.176***&1.15**&1.166**&1.108**&1.127**\\
				6&0.834**&0.986***&1.07***&1.131***&1.15***&1.287***&1.297***&1.27***&1.287***&1.228***&1.247***\\
				7&0.918***&1.07***&1.153***&1.215***&1.233***&1.370***&1.380***&1.354***&1.370***&1.312***&1.330***\\
				8&0.575**&0.728**&0.811**&0.873**&0.891**&1.028**&1.038***&1.011**&1.028**&0.969**&0.988**\\
				9&0.578**&0.73**&0.814**&0.875**&0.893**&1.031***&1.040***&1.014**&1.030**&0.972**&0.991**\\
				10&0.251&0.403*&0.487*&0.548*&0.567*&0.704**&0.714**&0.687*&0.704*&0.645&0.664\\
				11&0.115&0.267&0.351&0.412&0.431&0.568*&0.578*&0.551&0.568&0.509&0.528\\
				12&&0.152&0.236&0.297&0.316&0.453&0.463&0.436&0.453&0.394&0.413\\
				13&&&0.084&0.145&0.163&0.301&0.31&0.284&0.3&0.242&0.261\\
				14&&&&0.061&0.08&0.217&0.227&0.2&0.217&0.158&0.177\\
				15&&&&&0.018&0.156&0.165&0.139&0.155&0.097&0.116\\
				16&&&&&&0.137&0.147&0.121&0.137&0.079&0.097\\
				17&&&&&&&0.010&-0.017&0.000&-0.059&-0.040\\
				18&&&&&&&&-0.026&-0.010&-0.068&-0.050\\
				19&&&&&&&&&0.016&-0.042&-0.023\\
				20&&&&&&&&&&-0.058&-0.040\\
				EFF&&&&&&&&&&&0.019\\
				
				From/To & ANN  & 22 & 24 & 26 & 31 & 36&&&&&\\
				11&0.528&0.567&0.475&0.681&0.539&1.027&&&&&\\
				12&0.413&0.452&0.36&0.566&0.424&0.912&&&&&\\
				13&0.261&0.299&0.208&0.414&0.271&0.760&&&&&\\
				14&0.177&0.216&0.124&0.33&0.188&0.676&&&&&\\
				15&0.116&0.154&0.063&0.269&0.126&0.615&&&&&\\
				ANN&0.097&0.136&0.044&0.251&0.108&0.597&&&&&\\
				17&-0.040&-0.001&-0.093&0.113&-0.029&0.459&&&&&\\
				18&-0.050&-0.011&-0.103&0.104&-0.039&0.450&&&&&\\
				19&-0.023&0.015&-0.076&0.13&-0.013&0.476&&&&&\\
				20&-0.040&-0.001&-0.093&0.113&-0.029&0.459&&&&&\\
				EFF&0.019&0.057&-0.034&0.172&0.03&0.518&&&&&\\
				22&&0.039&-0.053&0.153&0.011&0.499&&&&&\\
				
				\hline
			\end{tabular}%
			\label{tab:cars}%
		\end{tiny}
	\end{center}
	\raggedright
	\footnotesize{Notes: Cumulative abnormal returns are calculated using the fitted values from the CAPM. Full Sample includes all firms that were not de-listed from the DJSI in any industry-year where one or more firms were de-listed. Base Sample restricts the full sample to include all listed firms and only those other firms with an asset holding at least 80\% of that of the lowest entrant in their industry-year. Significance given for a two-sample t-test of equality between joining and non-joining firms, with * - 5\%, ** - 1\%, *** - 0.1\%}
\end{table}

CARs, discussed in Table \ref{tab:carsp} recognise the trend in the stocks performance prior to the listing, they offer a measure of how listing creates deviation from that pre-announcement path. For the stated start and end dates we test whether the CARs of a pooled sample of listed firms are equal to that of a pooled sample of non-listed firms over the whole fourteen years of data. Positive values signify that the recently DJSI listed firms are offering higher CARs. From dates in the table range from day one of the treatment period through to the day after changes become effective, whilst the end dates range from day 11 to the final day of the treatment period. Full results are reported in the appendix. Panel (a) shows that there are some positive CARs for samples starting a week before the announcement. Between announcement date and the date that changes become effective, CARs are significant and positive. Herein an opportunity for investors to profit is found. Around the effective date there is a correction. CARs that start on day 20, one day prior to the effective date, show significant negative CARs. Throughout the post effective date range we see negative CARs but few others are significant. When we focus only on the base sample the only notable significance that remains is the positive return surrounding the announcement date. Understandably the magnitude of these gains is smaller, but their continued existence merits further investigation.

Turning to de-listing effects, Table \ref{tab:carsx} may be sat neatly in contrast to Table \ref{tab:cars} from the listing analysis. Immediate observations are the greater magnitudes of the CARs and the increased number of holding periods for which significance of the CARs is noted. Strong evidence of a pre-announcement effect is provided in those groups starting on days 7 to 9 of the treatment period and ending on days 9 to 12. This is far more pronounced than that seen in Table \ref{tab:cars}. That both effects are positive raises questions about the role of the pre-announcement effect; investors may be thinking that these firms would retain listing. Around the announcement date itself there are few significant effects but early gains quickly give way to negative CARs as the correction effect kicks in. Note here that the smaller magnitude of gains through the period before the effective date means the corrections are smaller than those observed in Table \ref{tab:cars}. 

For the Base Sample the comparison between listing and de-listing is more stark. Picking up those shares that are to be de-listed offers higher CARs over large time ranges, provided the purchase of the shares takes place at least a week before the announcement date. Waiting until the announcement date offers little difference and hence any investor looking to take advantage would need to correctly identify those firms who were to de-list. From a trading perspective obtaining de-listed firms in the immediate aftermath of the announcement offers the highest probability of success; such shares offer a premium, albeit an insignificant one, in the base sample too.

\subsection{OLS Regressions}
\label{sec:ols}

To understand better the extent to which factors lie behind the observed CAR patterns we regress the CARs observed over five sub-periods from Table \ref{tab:sumstat} on the listing dummy, size, profitability and leverage. We study $CAR_i[from,to]$ as the dependent variable, where this is either $CAR_i[-15,1]$, $CAR_i[-15,15]$, $CAR_i[-1,1]$, $CAR_i[0]$ and $CAR_i[0,10]$. Regression is performed following equation \eqref{eq:carm}:
\begin{align}
CAR_i[from,to] = \alpha+\beta_{DJSI}DJSI^{+}_{iy}+\mathbf{\theta} X_{iy}+\gamma_j+\psi_{y}+\epsilon_{iy} \label{eq:carm}
\end{align}
Here $X_{iy}$ is the set of firm level covariates, $DJSI^{+}_{iy}$ is a dummy equal to 1 if firm $i$ joins the DJSI in year $y$. $\mathbf{\beta}$ is a vector of coefficients on the firm controls. $\gamma_j$ introduces fixed effects for industries where firm $i$ is in industry $j$. These fixed effects are incorporated to capture unobserved heterogeneity between industries, enabling the model to include any factors which act solely upon that sector. For the de-listing case the dummy for firm $i$ leaving the DJSI in year $y$ is $DJSI^{-}_{iy}$. $\psi_y$ is the year fixed effect that is added to represent the variation in conditions over time, this includes those which would have been brought about during the GFC. Remaining error terms, $\epsilon_{iy}$, are assumed to have constant variance and an expected value of 0. To address questions about the best choice of covariates, or whether they should enter linearly, quadratically or otherwise, we follow \cite{acemoglu2016value} to allow each of the three controls to enter as linear, squared and cubic. Robustness checks have been performed using just the linear, and then the linear with quadratic effects.  

\begin{table}
	\begin{center}
		\begin{tiny}
		\caption{OLS Regressions for Cumulative Abnormal Returns and Firm Characteristics \label{tab:ols}}
		\begin{tabular}{l c c c c c c c c c c }
			\hline
			
			& \multicolumn{5}{l}{Full Sample} & \multicolumn{5}{l}{Base Sample}\\
			From & 1 & 1 & 16 & 16 & 21 & 1 & 1 & 16 & 16 & 21\\
			To &  16 & 21 & 21 & 36 & 36 & 16 & 21 & 21 & 36 & 36\\ 
			\hline
			\multicolumn{11}{l}{Panel A: Listing}\\
			$DJSI^{+}$                & $0.449$        & $0.762$        & $0.384$        & $0.904$        & $0.373$        & $0.357$       & $0.520$       & $0.250$      & $0.926$      & $0.579$     \\
			& $(0.606)$      & $(0.694)$      & $(0.369)$      & $(0.730)$      & $(0.640)$      & $(0.460)$     & $(0.531)$     & $(0.273)$    & $(0.558)$    & $(0.497)$   \\
			Size                    & $2.072^{**}$   & $2.110^{*}$    & $-0.107$       & $1.111$        & $0.711$        & $-8.258$      & $-10.822$     & $-1.328$     & $-9.784$     & $-8.702$    \\
			& $(0.720)$      & $(0.825)$      & $(0.439)$      & $(0.869)$      & $(0.762)$      & $(5.565)$     & $(6.420)$     & $(3.302)$    & $(6.747)$    & $(6.009)$   \\
			Size$^2$                   & $-0.242^{**}$  & $-0.269^{**}$  & $-0.009$       & $-0.178$       & $-0.110$       & $0.741$       & $0.988$       & $0.117$      & $0.786$      & $0.692$     \\
			& $(0.091)$      & $(0.104)$      & $(0.055)$      & $(0.109)$      & $(0.096)$      & $(0.533)$     & $(0.615)$     & $(0.316)$    & $(0.646)$    & $(0.575)$   \\
			Size$^3$                   & $0.009^{*}$    & $0.011^{*}$    & $0.001$        & $0.008$        & $0.005$        & $-0.021$      & $-0.028$      & $-0.002$     & $-0.018$     & $-0.016$    \\
			& $(0.004)$      & $(0.004)$      & $(0.002)$      & $(0.004)$      & $(0.004)$      & $(0.017)$     & $(0.019)$     & $(0.010)$    & $(0.020)$    & $(0.018)$   \\
			Profitability                     & $-0.799^{***}$ & $-1.538^{***}$ & $-0.460^{***}$ & $0.547^{*}$    & $1.039^{***}$  & $-1.540^{*}$  & $-2.355^{**}$ & $-0.508$     & $1.181$      & $1.470^{*}$ \\
			& $(0.221)$      & $(0.253)$      & $(0.135)$      & $(0.267)$      & $(0.234)$      & $(0.641)$     & $(0.739)$     & $(0.380)$    & $(0.777)$    & $(0.692)$   \\
			Profitability$^2$                    & $-0.025$       & $-0.056$       & $-0.018$       & $0.001$        & $-0.011$       & $0.223$       & $1.374$       & $1.046^{**}$ & $1.624^{*}$  & $0.642$     \\
			& $(0.031)$      & $(0.035)$      & $(0.019)$      & $(0.037)$      & $(0.032)$      & $(0.642)$     & $(0.741)$     & $(0.381)$    & $(0.778)$    & $(0.693)$   \\
			Profitability$^3$                 & $0.003$        & $0.005$        & $-0.002$       & $-0.006$       & $-0.008^{*}$   & $-0.010$      & $-0.201$      & $-0.182^{*}$ & $-0.362^{*}$ & $-0.183$    \\
			& $(0.003)$      & $(0.004)$      & $(0.002)$      & $(0.004)$      & $(0.003)$      & $(0.127)$     & $(0.147)$     & $(0.075)$    & $(0.154)$    & $(0.137)$   \\
			Leverage                     & $-3.095$       & $-3.405$       & $-0.475$       & $4.493$        & $4.041^{*}$    & $5.439$       & $6.659$       & $2.011$      & $2.487$      & $-0.548$    \\
			& $(1.926)$      & $(2.206)$      & $(1.173)$      & $(2.323)$      & $(2.036)$      & $(4.448)$     & $(5.132)$     & $(2.639)$    & $(5.393)$    & $(4.803)$   \\
			Leverage$^2$                    & $-1.317$       & $0.242$        & $1.790$        & $-16.203^{**}$ & $-14.976^{**}$ & $-22.080^{*}$ & $-25.538^{*}$ & $-5.290$     & $-10.315$    & $-3.274$    \\
			& $(5.105)$      & $(5.847)$      & $(3.110)$      & $(6.157)$      & $(5.397)$      & $(10.471)$    & $(12.081)$    & $(6.213)$    & $(12.696)$   & $(11.307)$  \\
			Leverage$^3$                    & $4.930$        & $3.692$        & $-1.432$       & $13.478^{**}$  & $12.460^{**}$  & $19.549^{**}$ & $20.665^{*}$  & $2.239$      & $5.560$      & $2.137$     \\
			& $(3.679)$      & $(4.215)$      & $(2.241)$      & $(4.437)$      & $(3.890)$      & $(7.137)$     & $(8.234)$     & $(4.235)$    & $(8.654)$    & $(7.707)$   \\
			\hline
			 
			R$^2$       & 0.044          & 0.036          & 0.030          & 0.032          & 0.041          & 0.034         & 0.039         & 0.051        & 0.089        & 0.083       \\
			Adj. R$^2$  & 0.042          & 0.034          & 0.028          & 0.030          & 0.039          & 0.022         & 0.028         & 0.040        & 0.078        & 0.072       \\
			\hline
			\multicolumn{11}{l}{Panel B: De-listing}\\
			$DJSI^{-}$ &7.260  & 5.709 & 2.762 & 0.496 & -0.889 & 2.643 & 1.031 & -2.947 & -0.122 & 1.544 \\
			&(5.240) & (4.661) & (8.982) & (5.270) & (6.223) & (3.788) & (3.324) & (6.170) & (3.856) & (4.588) \\
			Size&5.249 & 3.616 & -4.01 & -1.932 & -5.835 & 64.293 & 66.418 & 88.743 & -11.106 & -61.942 \\
			&(5.360) & -4.768 & -9.188 & -5.391 & -6.365 & -51.066 & -44.811 & -83.171 & -51.976 & -61.843 \\
			Size$^2$ & 0.691 & -0.622 & -0.012 & 0.064 & 0.635 & -5.890 & -6.048 & -7.787 & 1.137 & 5.688 \\
			&-0.667 & -0.594 & -1.144 & -0.671 & -0.792 & -4.747 & -4.166 & -7.732 & -4.832 & -5.749 \\
			Size$^3$ &0.026 & 0.027 & 0.013 & 0.003 & -0.02 & 0.179 & 0.182 & 0.225 & -0.038 & -0.172 \\
			&-0.027 & -0.024 & -0.046 & -0.027 & -0.032 & -0.145 & -0.128 & -0.237 & -0.148 & -0.176 \\
			Profitability & \multicolumn{1}{l}{-15.058***} & \multicolumn{1}{l}{-12.941***} & -3.496 & \multicolumn{1}{l}{2.683**} & \multicolumn{1}{l}{6.177***} & \multicolumn{1}{l}{-15.607***} & \multicolumn{1}{l}{-8.209**} & \multicolumn{1}{l}{19.354***} & \multicolumn{1}{l}{21.407***} & \multicolumn{1}{l}{25.412***} \\
			&-1.33 & -1.183 & -2.28 & -1.338 & -1.58 & -4.378 & -3.842 & -7.13 & -4.456 & -5.302 \\
			Profitability$^2$ & 0.126 & 0.097 & -0.28 & -0.03 & 0.062 & 5.223 & 0.241 & \multicolumn{1}{l}{-15.126*} & \multicolumn{1}{l}{-20.942***} & \multicolumn{1}{l}{-27.556***} \\
			&-0.362 & -0.322 & -0.62 & -0.364 & -0.43 & -5.09 & -4.467 & -8.29 & -5.181 & -6.164 \\
			Profitability$^3$ & \multicolumn{1}{l}{0.033*} & \multicolumn{1}{l}{0.03*} & -0.003 & 0.005 & 0.004 & -0.542 & 0.235 & 2.400   & \multicolumn{1}{l}{3.354***} & \multicolumn{1}{l}{4.521***} \\
			&-0.018 & -0.016 & -0.03 & -0.018 & -0.021 & -1.004 & -0.881 & -1.635 & -1.021 & -1.215 \\
			Leverage & \multicolumn{1}{l}{-33.445**} & -18.957 & 25.893 & \multicolumn{1}{l}{29.215**} & \multicolumn{1}{l}{35.903**} & -5.904 & 6.141 & 23.646 & -12.642 & -24.608 \\
			& -13.951 & -12.409 & -23.913 & -14.031 & -16.567 & -35.025 & -30.735 & -57.045 & -35.649 & -42.417 \\
			Leverage$^2$& 27.776 & 9.312 & -53.756 & \multicolumn{1}{l}{-83.903**} & \multicolumn{1}{l}{-98.819**} & -17.421 & -40.571 & -95.016 & -13.013 & 18.037 \\
			& -36.709 & -32.65 & -62.92 & -36.919 & -43.593 & -79.411 & -69.685 & -129.338 & -80.826 & -96.171 \\
			Leverage$^3$ & 7.497 & 12.28 & 31.797 & \multicolumn{1}{l}{62.713**} & \multicolumn{1}{l}{72.447**} & 24.627 & 37.366 & 76.436 & 20.12 & -2.318 \\
			&-26.407 & -23.487 & -45.262 & -26.558 & -31.359 & -53.008 & -46.516 & -86.335 & -53.953 & -64.196 \\
			\hline
			$R^2$ & 0.010  & 0.007 & 0.014 & 0.006 & 0.008 & 0.015 & 0.013 & 0.036 & 0.012 & 0.015 \\
			$Adj-R^2$ & 0.010  & 0.007 & 0.013 & 0.005 & 0.007 & 0.012 & 0.011 & 0.029 & 0.010  & 0.013 \\

			\hline
			\multicolumn{11}{l}{\scriptsize{$^{***}p<0.001$, $^{**}p<0.01$, $^*p<0.05$}}
		\end{tabular}
	\end{tiny}
	\end{center}
\raggedright
\footnotesize{Notes: Coefficients are reported for the base sample (All), and for the reduced sample with just firms with a size at least 80\% of the size of the smallest firm gaining listing (Large). Selected pairings of from and to dates are shown. Coefficients from regression $CAR_i[from,to] = \alpha+\beta_{DJSI}DJSI_{iy}+\mathbf{\theta}X_{iy}+\gamma_j+\psi_{y}+\epsilon_{iy}$ for cumulative abnormal returns between the from and to dates stated at the top of the column, $CAR_i[from,to]$. $DJSI_{y}$ is a dummy taking the value 1 if firm $i$ joins the DJSI in year $y$. $X_{iy}$ is a vector of common firm characteristics associated with stock returns, being size (log assets), return on equity and leverage (ratio of debt to capital). All characteristics are included as linear, quadratic and cubic. $\gamma_j$ is an industry fixed effect where firm $i$ is in industry $j$ as defined by the North American Industry Classification System at the two-digit level. $\psi_y$ are year fixed effects. Figures in parentheses report robust standard errors. Significance denoted by * = 1\%, ** = 5\%, and ***=1\%. $N$=22159}
\end{table}

Panel A shows that across all five periods, and for both samples, the main consistency observed is that the DJSI joining dummy is not significant in any of the ten equations. Such a result is opposite to the univariate tests of the previous section, but is entirely in line with the ambiguity of conclusions on listing effects in the current literature. Firm size is used to split the sample and for the full sample log assets has significant coefficients on the linear, quadratic and cubic terms. Only for the full 31 day period is no significance really seen. By contrast in the base sample very few of these size coefficients are significant. Profitability is significant in the linear term, but not for the quadratic or cubic. Leverage in these equations is also significant in the full sample, this can be linked to the correlations observed in Table \ref{tab:sumstat}. When reducing to the base sample much of the significance of leverage disappears. 

Panel B of Table \ref{tab:ols} indicates no significance to any of the de-listing dummies, this is consistent with the message on listing also. However, in the listing case there were significant effects arising from firm size; such are not found in the de-listing results. Consequently these tables offer little motivation for the movement to a base sample. Profitability coefficients become larger in magnitude in the base sample, whilst the significance of some leverage coefficients from the full sample disappear when only the larger firms have focus. Table \ref{tab:moreols} looks at an extended set of time ranges and reports only the coefficient on the DJSIX dummy. Occasional evidence of significance is seen. As in the listing analysis there is limited evidence of a DJSIX effect once firm characteristics are controlled for. Because the generalised synthetic control results look only at the full sample to maximise the possible candidates for the matched portfolio this lack of motivation for a reduced sample serves to aid the case for the \cite{xu2017generalized} approach adopted in this paper.

\begin{table}[htbp]
	\begin{center}
	\caption{Estimated Listing Effect from Cumulative Abnormal Returns OLS Regressions \label{tab:moreols}}
	\begin{tiny}
	\begin{tabular}{llrrrrrrrrrrr}
		\hline
		&\multicolumn{1}{l}{From} & 1     & 1     & 1     & 1     & 1     & 1     & 1     & 1     & 1     & 6     & 6 \\
		&\multicolumn{1}{l}{To} & 15     & ANN     & 17    & 18    & 19    & 20    & EFF    & 22 & 26 & 11 & 15 \\
		\hline
		$DJSI^{+}$ & Full & 0.378 & 0.449 & 0.379 & 0.484 & 0.834 & 0.909 & 0.762 & 0.642 & 0.850 & 0.377 & 0.357 \\
		&& (0.591) & (0.606) & (0.620) & (0.631) & (0.653) & (0.691) & (0.694) & (0.720) & (0.825) & (0.350) & (0.480)\\
		& Base & 0.270 & 0.357 & 0.383 & 0.517 & 0.663 & 0.618 & 0.520 & 0.398 & 0.729 & 0.334 & 0.286 \\
		&& (0.452) & (0.460) & (0.471) & (0.481) & (0.499) & (0.527) & (0.531) & (0.549) & (0.636) & (0.260) & (0.366)\\
		$DJSI^{-}$ & Full & 6.888 & 7.260 & 6.328 & 6.425 & 6.516 & 6.398 & 5.709 & 5.704 & 3.978 & 15.80** & 12.09***\\
		&& (5.430) & (5.244) & (5.061) & (4.932) & (4.799) & (4.783) & (4.661) & (4.554)& (4.244) & (8.006) & (6.029) \\
		& Base & 2.622 & 2.643 & 2.221 & 1.867 & 1.390 & 0.603 & 1.031 & 1.323 & 0.126 & 12.21** & 7.253\\
		&& (3.927) & (3.788) & (3.572) & (3.575) & (3.488) & (3.403) & (3.324) & (3.247) & (3.033) & (5.997) & (4.838)\\
		\hline
		&\multicolumn{1}{l}{From} & 6     & 6     & 6     & 6     & 6     & 11     & 11     & 11     & 11     & 11     & 11 \\
		&\multicolumn{1}{l}{To} & ANN     & 17     & 20    & EFF    & 26    & 11    & 15    & ANN & 17 & EFF & 26 \\
		\hline
		$DJSI^{+}$ & Full & 0.428 & 0.359 & 0.888 & 0.741 & 0.829 & 0.108 & 0.088 & 0.159 & 0.090 & 0.472 & 0.560 \\
		&& (0.496)& (0.513) & (0.591) & (0.596) & (0.728) & (0.145) & (0.351) & (0.371) & (0.389) & (0.496)& (0.632)\\
		& Base & 0.373 & 0.399 & 0.633 & 0.536 & 0.745 & 0.089 & 0.041 & 0.129 & 0.155 & 0.291 & 0.501 \\
		&& (0.375) & (0.385) & (0.436) & (0.444) & (0.546) & (0.114) & (0.284) & (0.294) & (0.305) & (0.381) & (0.489)\\
		$DJSI^{-}$& Full & 12.16* & 11.14* & 9.703* & 8.592 & 5.762 & 17.82 & 8.786 & 9.462 & 8.098 & 5.500 & 2.752 \\
		&& (6.528) & (6.198) & (5.611) & (5.487) & (4.834) & (19.34) &(10.64)&(9.468) & (8.578) & (6.911) & (5.716) \\
		& Base & 6.863 & 5.913 & 3.018 & 3.428 & 1.737 & 14.90 & 2.832 & 2.853 & 1.798 & -0.320 & -1.369 \\
		&& (4.586) & (4.380) & (3.918) & (3.829) & (3.400) & (14.08) & (7.059) & (6.355) & (5.850) & (4.678) & (3.949)  \\
		\hline
		&\multicolumn{1}{l}{From} & 15     & 15     & 15     & 15     & 15      & ANN     & ANN     & ANN     & ANN & ANN     & ANN \\
		&\multicolumn{1}{l}{To} & 15 & ANN    & 17     & 20    & EFF    & 26    & ANN & 17 & 18 & 19 & 20 \\
		\hline
		$DJSI^{+}$ & Full & 0.087 & 0.158 & 0.088 & 0.618 & 0.471 & 0.559 & 0.071 & 0.002 & 0.107 & 0.457 & 0.531\\
		&& (0.155) & (0.206) & (0.244) & (0.379) & (0.395) & (0.544) & (0.140) & (0.196) & (0.246) & (0.292) & (0.349) \\
		& Base & 0.113 & 0.201 & 0.227 & 0.461* & 0.363 & 0.573 & 0.088 & 0.114 & 0.248 & 0.393* & 0.348\\
		&& (0.119) & (0.151) & (0.178) & (0.275) & (0.290) & (0.417) & (0.100) & (0.142) & (0.182) & (0.213) & (0.256)\\
		$DJSI^{-}$ & Full & -12.97 & -0.061 & -0.069 & 1.946 & 0.515 & -1.073 & 12.84 & 6.379 & 5.101 & 5.119 & 4.928 \\
		&& (21.21) & (14.62) & (11.68) & (5.947) & (3.728) & (3.706) & (19.16) & (13.54) & (15.52) & (10.11) & (9.570)\\
		& Base & -8.415 & -2.727 & -3.329 & -5.947 & -3.728 & -3.706 & 2.961 & -0.786 & -1.989 & -3.231 & 5.454 \\
		&& (14.30) & (10.22) & (8.547) & (6.230) & (5.781) & (4.553) & (13.27) & (10.04) & (8.482) & (7.466) & (5.718) \\
		\hline
			&\multicolumn{1}{l}{From} &  ANN     & ANN  & ANN     & 17    & 17     & 18     & 19  & 20 & 20 & EFF & EFF  \\
		&\multicolumn{1}{l}{To} & EFF    & 22 & 26    & 17 & EFF & EFF & EFF & EFF & 22& EFF & 22  \\
		\hline
		$DJSI^{+}$ & Full & 0.384 & 0.264 & 0.472 & -0.069 & 0.313 & 0.382 & 0.277 & -0.073 & -0.192 & -0.147 & -0.267 \\
		&& (0.369) & (0.399) & (0.530) & (0.133) & (0.334) & (0.307) & (0.265) & (0.218) & (0.261) & (0.162) & (0.218)  \\
		& Base & 0.250 & 0.129 & 0.460 & 0.026 & 0.162 & 0.136 & 0.002 & -0.143 & -0.265 & -0.098 & -0.219\\
		&& (0.273) & (0.300) & (0.409) & (0.098) & (0.248) & (0.223) & (0.193) & (0.154) & (0.190) & (0.122) & (0.166) \\
		$DJSI^{-}$ & Full & 2.762 & 3.166 & 0.909 & -0.085 & 0.745 & 0.953 & 1.416 & -1.954 & -8.073 & 6.564 & -1.242     \\
		&& (8.982) & (8.276) & (6.736) & (18.44) & (10.03) & (11.60) & (13.65) & (17.53) & (13.70) & (23.44) & (15.91) \\
		& Base & -2.947 & -1.460 & -3.278 & -4.532 & -4.128 & -4.027 & -3.985 & -2.376 & 0.900 & 9.588 & 8.522 \\
		&& (6.170) & (5.765) & (4.735) & (13.84) & (6.847) & (7.696) & (8.820) & (10.75) & (8.089) & (14.53) & (10.40) \\  	
		\hline
		&\multicolumn{1}{l}{From} &  EFF     & EFF  & EFF     & 22    & 22     & 22     & 26  & 26 & 31 &31 & 36   \\
		&\multicolumn{1}{l}{To} & 26    & 31 & 36    & 22 & 26 & 36 & 26 & 36 & 31 & 36 & 36 \\
		\hline
		$DJSI^{+}$ & Full & -0.059 & -0.158 & 0.373 & -0.120 & 0.088 & 0.520 & -0.060 & 0.372 & -0.144 & 0.387 & -0.010\\
		&&  (0.393) & (0.516) & (0.640) & (0.151) & (0.363) & (0.615) & (0.159) & (0.517) & (0.168) & (0.388) & (0.153)\\
		& Base & 0.112 & 0.135 & 0.579 & -0.121 & 0.210 & 0.676 & -0.067 & 0.400 & 0.004 & 0.447 & 0.048 \\
		&& (0.300) & (0.402) & (0.497) & (0.110) & (0.282) & (0.476) & (0.129) &(0.408) & (0.148) & (0.317) & (0.128)\\
		$DJSI^{-}$ & Full & -4.091 & -3.263 & -0.880 & 5.590 & -3.295 & -0.410 & 5.766 & 1.463 & -25.81 & -0.705 & -20.40 \\
		&&  (9.329) & (7.684) &(6.223) & (20.71) & (10.08) & (6.434) & (21.57) & (7.729) & (22.41) & (9.605) & (21.39) \\
		& Base & -1.464 & 0.297 & 1.544 & 7.457 & -3.685 & 1.008 & -5.408 & 2.553 & -8.292 & 2.191 & -14.74\\
		&&  (6.568) & (5.491) & (4.588) & (13.59) & (7.240) & (4.783) & (16.05) & (5.816) & (16.16) & (7.492) & (16.64)  \\  	
		\hline
	\end{tabular}
\end{tiny}
\end{center}
\raggedright
\footnotesize{Notes: Coefficients on gaining DJSI listing are reported for the base sample (Full), and for the reduced sample with just firms with a size at least 80\% of the size of the smallest firm gaining listing (Base). Coefficients for listing are from regression $CAR_i[from,to] = \alpha+\beta_{DJSI}DJSI^{+}_{iy}+\mathbf{\theta}X_{iy}+\gamma_j+\psi_{y}+\epsilon_{iy}$ and are estimated for cumulative abnormal returns between the stated from and to dates, $CAR_i[from,to]$. $DJSI^{+}_{iy}$ is a dummy taking the value 1 if firm $i$ joins the DJSI in year $y$. $X_{iy}$ is a vector of common firm characteristics associated with stock returns, being size (log assets), return on equity and leverage (ratio of debt to capital). All characteristics are included as linear, quadratic and cubic. $\gamma_j$ is an industry fixed effect where firm $i$ is in industry $j$ as defined by the North American Industry Classification System at the two-digit level. $\psi_y$ are year fixed effects. For de-listing regressions are repeated but with the de-listing dummy, $DJSI^{-}_iy$ replacing $DJSI^{+}_iy$. Figures in parentheses report robust standard errors. Significance denoted by * = 1\%, ** = 5\%, and ***=1\%.}
\end{table}

Table \ref{tab:moreols} reports a wider set of ranges for the CAR, providing coefficients on the DJSI joining dummy. These models maintain the full set of controls and fixed effects from Table \ref{tab:ols}, but the full details are not reported for brevity. There are now some significant coefficients at the 10\% level, but these account for less than 5\% of all the coefficients reported. As such this extended set does little to reverse the conclusions of a lack of DJSI joining abnormal return that was seen in Table \ref{tab:ols}. 

Regressions presented here suggest that much of the difference assigned to a new DJSI listing by the two-sample tests may actually be a consequence of other characteristics. Attributing effects to the correct characteristic represents one of the many challenges of using a testing approach. 

\section{Generalised Synthetic Control Approach}
\label{sec:synth}

Synthetic control methodologies \citep{abadie2003economic,abadie2010synthetic} seek to construct a counterfactual for a treated unit under the assumption that the treatment was not administered. Inherently unobservable the counterfactual is used purely for identifying the treatment effect, being the difference between the observed unit and the observed unit's counterfactual. In the assessment of excess stock returns from DJSI listing, the unit is the firm that gains listing and the treatment is the listing. This paper departs from the \cite{abadie2003economic} family of models by introducing the generalised approach of \cite{xu2017generalized}. Departure here owes to the fact that in many instances there are multiple firms gaining listing within the same industry and there is a strong likelihood of co-integrating relationships amongst stocks. Before presenting the results an outline of the \cite{xu2017generalized} approach is provided.

\subsection{Methodology}
%\label{sec:method}

Following \cite{acemoglu2016value} we construct a portfolio from the other stocks within the firms industry, selected using the two digit NAICS code. As discussed the portfolio is assembled using observations from the first trading date in November of the previous year, to 16 trading days ahead of the formal listing announcement. This typically provides a training set of 230 days\footnote{Because of the annual cycle of the DJSI listings we do not include a full year of training data.}. The synthetic control is then analysed for the period between three weeks in advance of the listing announcement and three weeks after the announcement. In total this gives a 31 trading day long period. This period reconciles with the observations of past event studies that impacts die out soon after listing but allows for the discovery of new effects in the post-listing that mirror identified time spans from pre-listing.  

In any given industry firms are split into a treatment group, $\mathcal{T}$, and a control group, $\mathcal{C}$. Treated firms are those who gain listing to the DJSI and the controls are all other firms in the same NAICS 2-digit code. Of the $N$ firms with sufficient data in a year, $N_{tr}$ are treated and the remaining $N_{co}$ are controls, such that $N_{co}+N_{tr}=N$. Each firm, $i$, is observed for $T$ periods, covering the $T_{0,i}$ control periods prior to listing, and the $q_i = T - T_{0,i}$ evaluation periods following the listing. $Y_{tr}$. The \cite{xu2017generalized} approach offers the possibility of differing numbers of observations for each firm. However for simplicity this exposition has $T_{0,i} = T_0$ and $q_i=q$. It is thus assumed that the outcome of interest, excess returns for firm $i$ at time $t$, $R_{it}$, are given by a linear factor model, equation \eqref{eq:lfm1}.
\begin{align}
R_{it}=\delta_{it}D_{it}+x_{it}' \beta+\lambda_i'f_t+\epsilon_{it} \label{eq:lfm1}
\end{align}
The treatment dummy, $D_{it}$, takes the value 1 for firms obtaining listing on the DJSI, that is $i \in \mathcal{T}$ and $t>T_0$. For our purposes there are no controls and so we can simplify the exposition to remove $x_{it}' \beta$. 

Innovation in \cite{xu2017generalized} draws on the $\lambda_i'f_t$ factor model which expands to \eqref{eq:lfm2}:
\begin{align}
\lambda_i'f_t=\lambda_{i1}f_{1t}+\lambda_{i2}f_{2t}+...+\lambda_{ir}f_{rt} \label{eq:lfm2}
\end{align}
This takes a linear additive form that covers conventional additive unit and time fixed effects as special cases. Many further common financial models are also permissable, including autoregressive components\footnote{See discussion in \cite{xu2017generalized} and \cite{gobillon2016regional}.}.

Define $R_{it}(1)$ as the excess stock return for firm $i$ at time $t>T_0$ and $R_{it}(0)$ as the pre-treatment excess returns for firm $i$ at time $t\leq T_0$. $\delta_{it}=R_{it}(1)-R_{it}(0)$ for any $i \in \mathcal{T}$, $t>T_0$. It may be written that:
\begin{align}
R_i=D_i\odot\delta_i+F\lambda_i+\epsilon_i, i \in 1,2,...,N_{co},N_{co}+1,N
\end{align}
in which $R_i=|R_{i1},R_{i2},...,R_{iT}|$;$D_i=|D_{i1},D_{i2},...,D_{iT}|'$, $\delta_i=|\delta_{i1},\delta_{i2},...,\delta_{iT}|$, and $\epsilon_i=|\epsilon_{i1},\epsilon_{i2},...,\epsilon_{iT}|'$ are $T\times 1$ vectors. The factors $F=|f_1,f_2,...,f_T|'$ is a $(T \times r)$ matrix. Determination of $r$ is discussed subsequently.

Stacking all $N_{co}$ control units together produces $R_{co}=[R_1,R_2,...,R_{N_{co}}]$ and $\epsilon_{co}=[\epsilon_1,\epsilon_2,...,\epsilon_{N_co}]$, the factor matrix, $\Lambda_{co}=[\lambda_1,\lambda_2,...,\lambda_{N_{co}}]$, is $(N_{co}\times r)$, whilst $R_{co}$ and $\epsilon_{co}$ are both $(T \times N_{co})$. The stacked model is stated as equation \eqref{eq:lfm3}:
\begin{align}
R_{co}=F\Lambda_{co}'+\epsilon_{co} \label{eq:lfm3}
\end{align}  
Identification of the parameters is constrained by a requirement that $F'F/T=I_r$ and $\Lambda_{co}'\Lambda_{co}=$diagonal. Average listing effects for those who are listed on the DJSI, are then the average effects of treatment on the treated $(ATT)$. At time $t$, $t>T_0$ the ATT, $ATT_{t,t>T_0}$ is estimated as per equation \eqref{eq:lfm4}\footnote{For more on the social economic interpretation of this see \cite{blackwell2018make}.}.
\begin{align}
ATT_{t,t>T_0}=1/{N_{tr}}\sum_{i\in\tau}^{}[Y_{it}(1)-Y_{it}(0)]=\dfrac{1}{N_{tr}}\sum_{i\in\tau}\delta_{it} \label{eq:lfm4}
\end{align}

\cite{xu2017generalized}, like \cite{abadie2010synthetic}, treat the treatment effects $\delta_{it}$ as conditional on the sample data. Identification of these necessitates an appropriate measure of $R_{it}(0)$ when $t>T_0$ and $i \in \mathcal{T}$\footnote{A discussion of the requirements for causal inference in the generalised synthetic control framework is provided as a supplementary appendix to \cite{xu2017generalized}.}. Estimation of the parameters of the model proceeds using three steps. Firstly estimates for $\hat{F}\hat{\lambda}_co$ are obtained through:
\begin{align}
(\hat{F},\hat{\Lambda_{co}})=\underset{\tilde{\beta},\tilde{F}\tilde{\Lambda}_{co}}{argmin}\sum_{i\in \mathcal{C}}(R_i-\tilde{F}\tilde{\Lambda}_i)'(R_i-\tilde{F}\tilde{\Lambda}_i) \label{eq:lfm5} 
\end{align}

Recalling that this minimisation is performed subject to the twin constraints that $\tilde{F}'\tilde{F}/T=I_r $ and that $\tilde{\Lambda}_{co}'\tilde{\Lambda}_{co}$ is a diagonal matrix.

Following \cite{xu2017generalized} the factor loadings are calculated. Values restricted to the pre-announcement period gain subscript ``0'''s. Hats denote estimates from \eqref{eq:lfm5}. Step 2 is thus:
\begin{align}
\hat{\lambda}_i&=\underset{\hat{\lambda_i}}{argmin}(R_i^0-\hat{F}^0\tilde{\lambda_i})'(R_i^0-\hat{F}^0\tilde{\lambda_i})\label{eq:lfm6}\\
&={(\hat{F}^{0\prime}\hat{F}^0)}^{-1}\hat{F}^{0\prime}R_i^0, i\in \mathcal{T} \nonumber
\end{align}
Step 3 calculates treated counterfactuals based on the estimated $\hat{F}$ and $\hat{\lambda_{co}}$:
\begin{align}
\hat{R}_{it}(0)=\hat{\lambda_i}'\hat{f}_t, i \in \mathcal{T}, t>T_0
\end{align}
Estimates for the average treatment effect on the treated, $ATT_t$ are provided as:
\begin{align}
\hat{ATT}_t=(1/N_{tr})\sum_{i \in \mathcal{T}}[R_{it}(1)-\hat{R}_{it}(0)]\; \text{for}\; t>T_0
\end{align}
In order to obtain convergence in the estimated factor loadings it is required that there be sufficiently large numbers of controls and a sufficiently long control period. As we have more than 200 days of data, and a large number of firms in each two digit NAICS code, there would not be expected to be any concerns about convergence. Indeed in every case the reported tests of convergence reveal that the model does converge.

Within the \cite{xu2017generalized} the number of factors to be included is determined using a five step procedure. Firstly a given $r$ is selected and an interactive fixed effects (IFE) model is estimated for the control group data to obtain an estimate of $F$, $\hat{F}$. A cross-validation loop is run at step 2 which first works systematically through the control period omitting one period and obtaining factor loadings for each treated unit, $i$, according to the formula:
\begin{align}
\hat{\lambda}_{i,-s}=={(F_{-s}^{0\prime} F_{-s}^0)}^{-1}F_{-s}^{0\prime}R_{i-s}^0, i\in \mathcal{T}
\end{align}
where the use of $-s$ in the subscripts denotes the ommision of period $s$ from the estimation. Predicted outcomes for the missing period are saved and compared with the observed period $s$ return to construct a prediction error $e_{is}=R_{is}(0)-\hat{R}_{is}(0)$. Step 3 sees the calculation of the mean square predicted error (MSPE) given the selected number of factors. Given $r$ the MSPE is:
\begin{align}
MSPE(r)=\sum_{s=1}^{T_0}\sum_{i \in \mathcal{T}}e_{is}^2/T_0 \label{eq:lfm10}
\end{align}
Repeating the process over further possible $r$ enables the identification of $r^*$ as that number of factors which minimises the MSPE from equation \eqref{eq:lfm10}. \cite{xu2017generalized} demonstrate through Monte Carlo simulation that this simplistic procedure performs well in factor number selection\footnote{It is also shown to perform well in small datasets, but this is not a concern for our daily financial data \citep{xu2017generalized}.}. 

In order to obtain inference from the estimated ATT we need a conditional variance of the ATT estimator, i.e. $Var_{\epsilon}(\hat{ATT}_t|D_t,\Lambda F)$. Although $\epsilon$ should be the only random variable not being conditioned upon, it may be correlated with $\hat{\lambda}_i$ from the estimation loop above. Nonetheless, $\epsilon$ remains a measurement of the variations in returns that we cannot explain and which is unrelated to treatment assignment.

In an approach similar to \cite{acemoglu2016value}, \cite{xu2017generalized} proposes a four step algorithm for determining the uncertainty estimates of, and hence confidence intervals for, $\hat{ATT}_t$. Treated counterfactuals are simulated for control units, that is firms whose DJSI membership status does not change in the given year. For this purpose the resampling scheme is given as:
\begin{align*}
\tilde{R_i}(0)=\hat{F}\hat{\lambda_i}+\tilde{\epsilon}_i, && \forall i \in \mathcal{C}; \\
\tilde{R_i}(0)=\hat{F}\hat{\lambda_i}+\tilde{\epsilon}_i^p, && \forall i \in \mathcal{T}; \\
\end{align*}
where simulated outcomes from the event of the treatment not occurring are collected in $\tilde{R_i}(0)$, the estimated conditional mean is captured through the estimated factors, $\hat{F}\hat{\lambda_i}$ and the resampled residuals are incorporated through $\tilde{\epsilon}_i$ and $\tilde{\epsilon}_i^p$. The variance of the latter is liable to be bigger than the former since  $\hat{F}\hat{\lambda_i}$ is estimated from the control units and will therefore be expected to fit better on those firms that did not gain listing.

Step one is to start a loop that will run a specified number of times to generate a sufficiently large number of comparison observations for the confidence intervals. Within this element of the process it is necessary take a control unit, $i$, and act as if it has been treated in the time $t>T_0$. The rest of the control units are resampled with replacement to form a new sample which contains that new ``treated'' unit and a full set of $N_{co}$ controls. The generalized synthetic control method is applied to obtain a vector of residuals, $\hat{\epsilon}_{(m)}^p=R_i-\hat{R_i}(0)$. Collecting these residuals from every loop then creates a vector \textbf{$e^p$}. Step two applies the generalised synthetic control method to the original data to obtain the fitted average treatment effects, $\hat{ATT}_t$ for all time periods. Estimate coefficients and obtain fitted values and residuals for the control units, $\mathbf{\hat{R}_{co}}=\lbrace\hat{R}_1(0),\hat{R}_2(0),....\hat{R}_{N_{co}}(0)\rbrace$ and $\mathbf{\hat{e}}=\lbrace\hat{\epsilon}_1,\hat{\epsilon}_2,\hat{\epsilon_{N_{co}}}\rbrace$

Step three of the process then involves another bootstrap loop, operating for $B_2$ repetitions. For each repetition a bootstrapped sample, $S^{(k)}$  is used.  In this case in round $k\in\lbrace1,...,B_2$ the previous estimates of $\tilde{\epsilon}_i$ and $\tilde{\epsilon}_j^p$ are randomly selected from the sets $\mathbb{e}$ and $\mathbf{e^p}$. We fit $\hat{R}_i=\hat{F}\hat{\lambda}_i$ and hence construct a sample by:
\begin{align*}
\tilde{R_i}^{(k)}(0)=\hat{R}_i(0)+\tilde{\epsilon}_i & \i \in \mathcal{C} \\
\tilde{R_i}^{(k)}(0)=\hat{R}_j(0)+\tilde{\epsilon}_i^p & \j \in \mathcal{T} 
\end{align*}  
In this case \cite{xu2017generalized} notes that the simulated treatment counterfactuals do not contain the treatment effect. From here the generalised synthetic control is applied to the bootstrapped sampled $S^{(k)}$ new estimates for the average treatment effects, $\hat{ATT}_{t,t>T_0}$. Adding this estimate to the others creates a set of stored estimates allowing the final obtaining of the bootstrap estimator $\hat{ATT}^k_{t,t>T_0}$. Finally the variance of all of these collected average treatment effects may be calculated and the confidence intervals constructed accordingly. Advantageously this is implemented automatically within \cite{gsynth}. 

Contrast with the simpler approaches used in Section \ref{sec:first} comes from the calculation of the synthetic portfolio data, compared to the CAPM abnormal return estimates embeds a role for non-listed firms. In so doing the generalised synthetic control approach addresses one of the concerns raised by \cite{hawn2018investors}. Earlier t-test comparisons and dummy variables for listings both saw influences from other firms on the magnitude of the listing effect. As reported, the introduction of a further dependence on the relationship between the control and treatment groups embeds robustness to any shock which equally effects all in the industry. Further, by isolating those movements that would have happened under the assumption of an ongoing relationship with non-listed stocks, deviations can much more readily be attributed to the listing.

\subsection{Results}
\label{sec:results}

Cumulative abnormal returns are available for periods of one day, or longer, within the 36 day treatment period. A total of 630 combinations of start and end time are possible. Brevity dictates that only a selection of these may be reported, full results being provided in a supplementary appendix. Building on the precedent in \cite{acemoglu2016value,acemoglu2017power} and \cite{chamon2017fx}, we employ the generalised framework of \cite{xu2017generalized} to estimate said. This leap from the original \cite{abadie2003economic} and \cite{abadie2010synthetic} approach is taken because there are too often more than one company obtaining listing on the DJSI from any given industry. The original synthetic control cannot deal efficiently with such. Table \ref{tab:tandc} already highlighted the presence of multiple treatments within the industry-year. These figures are repeated within the fit comparisons of Table \ref{tab:tbi2}. 

Table \ref{tab:tbi2} primary purpose is to report the fit statistics for the generalised synthetic control and to offer comparison with the CAPM generated fits. These are reported for the in-sample control period. MSPE values are reported at the two digit NAICS code level to indicate the quality of the fit through the training period. In the majority of cases these values are below 2, with high values appearing only where the number of controls is low. There are many occasions near the GFC where the synthetic control model has an MSPE well below that associated with the CAPM, sometimes being less than half that of the original approach. Industry 21 in 2005 is a good example of this. In more recent years the number of times where the CAPM delivers a better fit is almost identical, though often the margin is very small. There remain times where the synthetic control error is less than half that of the CAPM model, including industry 33 in 2018 where the MSPE is just 1.424 compared to an MSPE of 4.102 for the CAPM. Overall there are 55 cases where the CAPM can be considered better fitting during the control period, compared to 82 for the generalised synthetic control approach. A t-test to compare the MSPE for the two models weighting all industry-years equally confirms a better fit from the generalised synthetic approach significant at the 5\% level.

\begin{table}
	\begin{center}
		\begin{tiny}
			\caption{Fit Statistics by Industry \label{tab:tbi2}}
			\begin{tabular}{l c c c c c c c c l c c c c c c c}
				\hline
				Year & NAICS2 & V. & Co. & Tr & Ctrl & \multicolumn{2}{l}{MSPE}&& Year & NAICS2 & V. & Co. & Tr & Ctrl & \multicolumn{2}{l}{MSPE}\\
				& & &&& & CAPM & Synth& &&    & &  & & & CAPM & Synth\\
				\hline
					
					2005&21&0&3&1&129&3.344&1.358&&2011&53&0&5&1&178&1.919&1.142\\
					2005&22&0&5&6&99&1.488&0.631&&2011&56&0&1&1&49&1.882&1.666\\
					2005&23&0&3&1&31&2.703&2.639&&2011&72&0&1&1&47&0.914&0.983\\
					2005&31&0&1&3&105&0.783&0.839&&2012&22&0&5&1&98&0.895&0.568\\
					2005&32&0&5&8&318&1.041&1.003&&2012&31&0&4&2&107&0.606&0.600\\
					2005&33&0&2&9&661&1.602&1.54&&2012&32&0&1&2&275&0.517&0.619\\
					2005&44&0&1&4&89&1.628&1.607&&2012&33&0&2&2&574&1.416&1.403\\
					2005&45&0&2&1&53&1.149&1.032&&2012&45&0&1&1&36&7.648&7.772\\
					2005&48&0&5&1&89&0.915&0.825&&2012&51&0&2&3&263&2.309&2.181\\
					2005&51&0&1&4&285&0.866&0.96&&2012&52&0&3&1&499&2.421&1.967\\
					2005&52&0&2&7&535&0.707&0.716&&2012&54&0&1&1&107&1.864&1.964\\
					2005&54&0&4&2&112&1.744&1.751&&2012&56&0&1&1&48&0.738&0.789\\
					2005&56&0&3&1&57&0.862&0.897&&2013&21&0&3&2&152&1.594&1.396\\
					2005&72&0&5&1&62&1.18&1.093&&2013&31&0&3&2&112&0.574&0.86\\
					2006&21&0&5&2&140&4.53&1.608&&2013&32&0&2&4&293&1.329&1.377\\
					2006&22&0&5&4&104&0.786&0.936&&2013&33&0&1&4&587&0.728&1.265\\
					2006&31&0&4&1&109&0.402&0.416&&2013&44&0&5&2&83&0.852&0.883\\
					2006&32&0&3&3&327&2.244&1.721&&2013&51&0&1&2&287&3.11&3.061\\
					2006&33&0&1&1&683&1.767&1.826&&2013&52&0&3&2&542&1.229&1.057\\
					2006&42&0&5&1&82&0.852&0.896&&2013&53&0&5&3&191&1.443&1.122\\
					2006&51&0&1&1&301&4.355&4.331&&2013&54&0&1&1&103&4.532&4.533\\
					2006&52&0&3&3&540&0.872&0.841&&2014&22&0&5&2&100&0.714&0.749\\
					2007&21&0&3&1&162&1.623&0.936&&2014&23&0&4&2&46&0.683&0.734\\
					2007&22&0&1&1&106&1.143&1.111&&2014&32&0&4&1&349&1.991&2.172\\
					2007&33&0&4&2&694&3.035&2.754&&2014&33&0&3&3&637&1.353&1.416\\
					2007&45&0&4&1&49&2.592&2.509&&2014&48&0&3&1&119&0.678&0.663\\
					2007&51&0&4&1&318&2.442&2.465&&2014&51&0&4&2&322&2.277&2.221\\
					2007&52&0&4&3&517&1.047&1.042&&2014&52&0&4&2&591&0.836&0.820\\
					2007&54&0&1&1&114&1.681&1.719&&2014&53&0&3&1&220&1.341&1.343\\
					2007&62&0&3&1&52&3.394&3.211&&2014&72&0&1&1&58&0.938&0.968\\
					2008&31&0&5&1&104&1.820&1.835&&2015&23&0&4&1&43&2.23&0.866\\
					2008&32&0&4&1&282&3.700&3.610&&2015&31&0&5&2&105&0.473&1.056\\
					2008&33&0&4&2&573&5.592&3.728&&2015&32&0&2&3&392&0.813&0.851\\
					2008&45&0&2&1&34&5.020&2.519&&2015&33&0&5&1&599&0.839&0.912\\
					2008&51&0&1&1&256&4.680&4.631&&2015&51&0&4&1&333&0.728&0.739\\
					2008&52&0&3&2&440&4.913&3.562&&2015&52&0&5&1&592&0.426&0.414\\
					2008&53&0&2&2&156&3.518&2.810&&2015&53&0&5&2&217&1.459&0.961\\
					2008&56&0&2&1&51&1.604&1.598&&2015&72&0&5&1&61&3.01&0.982\\
					2009&21&0&1&1&119&10.00&5.781&&2016&31&0&4&2&105&1.034&1.040\\
					2009&31&0&5&1&88&2.317&2.103&&2016&32&0&1&1&381&1.122&1.353\\
					2009&32&0&3&4&221&4.63&4.636&&2016&33&0&3&2&560&0.729&2.150\\
					2009&33&0&5&2&430&3.386&3.409&&2016&44&0&2&1&81&1.577&1.225\\
					2009&42&0&1&2&60&4.482&4.687&&2016&51&0&5&1&329&2.387&2.423\\
					2009&44&0&4&1&60&4.626&4.573&&2016&52&0&5&3&564&2.06&1.431\\
					2009&45&0&5&1&26&9.064&5.498&&2016&53&0&5&2&223&3.039&2.452\\
					2009&48&0&4&1&84&3.057&2.609&&2016&56&0&3&1&53&0.585&0.621\\
					2009&51&0&5&1&204&2.138&2.161&&2017&21&0&5&2&121&3.634&3.520\\
					2009&52&0&5&3&353&9.198&7.351&&2017&31&0&3&2&106&1.029&0.783\\
					2009&54&0&1&1&87&4.522&4.543&&2017&32&0&3&2&388&1.922&1.992\\
					2010&21&0&2&3&141&4.235&1.947&&2017&33&0&5&3&576&1.834&1.904\\
					2010&32&0&1&4&274&1.314&1.333&&2017&48&0&5&1&115&1.803&1.452\\
					2010&33&0&5&3&572&1.365&1.262&&2017&51&0&3&2&330&0.793&0.828\\
					2010&51&0&5&1&248&0.593&0.664&&2017&52&0&5&2&608&0.801&0.955\\
					2010&54&0&5&2&111&2.241&2.227&&2017&53&0&5&2&224&1.639&1.088\\
					2010&56&0&1&1&46&2.192&2.343&&2017&54&0&1&1&88&1.566&1.689\\
					2011&32&0&4&1&273&0.954&1.003&&2017&72&0&2&2&62&0.713&0.728\\
					2011&33&0&3&4&577&1.463&1.477&&2018&21&0&4&1&103&1.839&0.827\\
					2011&44&0&1&1&80&3.391&3.39&&2018&32&0&5&2&221&1.748&1.787\\
					2011&45&0&5&1&33&1.628&1.382&&2018&33&0&5&2&401&4.102&3.693\\
					2011&48&0&4&2&88&3.359&2.664&&2018&52&0&2&2&429&1.097&1.13\\
					2011&52&0&4&3&462&1.284&0.969&&2018&56&0&4&1&28&0.698&0.745\\
				\hline	
				
			\end{tabular}
		\end{tiny}
	\end{center}
\raggedright
\footnotesize{Notes: Models are fitted using the generalised synthetic control method of \cite{xu2017generalized}. NAICS2 reports the two-digit North American Industry Classification System (NAICS) code for the considered industry. MSPE is the Mean Squared Prediction Error when fitting the synthetic versions of the fitted shares to the training data. CAPM reports the MSPE for the CAPM based CARs from Section \ref{sec:simpcar}, whilst Synth reports the MSPE for the generalised synthetic control methodology. V. reports a test for the cointegration of the error matrix with 0 implying rejection. Co. gives the number of cointegrating relationships used in the construction of the unobserved parameter. Tr is the number of firms who joined the DJSI for that two digit NAICS code. Ctrl is the number of control firms used to construct the couterfactual model for entering firms. All firms with missing data are eliminated, including some new listings to the DJSI. All estimations performed using \textit{gsynth} \citep{gsynth}}
\end{table}

Model fit from the generalised synthetic control is again better than that from the corresponding CAPM, with the in sample MSPE comparison showing it to be the better fit in 31 cases compared to 24 for the CAPM. Table \ref{tab:exitrmse} provides the full comparison. This is a smaller differential than for the entering firms. As with entry where the generalised synthetic control does improve fit the margin of improvement is much larger, industry 51 in 2006 has two firms leaving the market and a MSPE of 4.355 from the CAPM but just 1.346 for the generalised synthetic control. Another parallel is seen in the more even performance of the two techniques in recent years. From Table \ref{tab:exitrmse} the number of cointegrating relationships in each model can be seen.    

\begin{table}
	\begin{center}
		\begin{tiny}
			\caption{Fit Statistics by Industry \label{tab:exitrmse}}
			\begin{tabular}{l c c c c c c c c l c c c c c c c}
				\hline
				Year & NAICS2 & V. & Co. & Tr & Ctrl & \multicolumn{2}{l}{MSPE}&& Year & NAICS2 & V. & Co. & Tr & Ctrl & \multicolumn{2}{l}{MSPE}\\
				& & &&& & CAPM & Synth& &&    & &  & & & CAPM & Synth\\
				\hline
				
				2006&22&0&2&2&106&0.786&0.895&&2013&33&0&4&3&588&0.728&0.895\\
				2006&31&0&4&1&109&0.402&0.846&&2013&44&0&4&1&84&0.852&1.923\\
				2006&33&0&1&2&682&1.767&1.521&&2013&52&0&3&3&541&1.229&3.065\\
				2006&51&0&5&2&300&4.355&1.346&&2014&32&0&5&1&349&1.991&0.679\\
				2006&52&0&4&2&541&0.872&1.044&&2014&51&0&2&1&323&2.277&0.719\\
				2007&51&0&5&1&318&2.442&2.307&&2014&72&0&1&1&58&0.938&0.417\\
				2007&52&0&4&2&518&1.047&0.911&&2015&23&0&3&1&43&2.23&1.137\\
				2008&32&0&4&1&282&3.7&2.917&&2015&31&0&4&1&106&0.473&0.398\\
				2008&33&0&4&1&574&5.592&2.841&&2015&33&0&5&4&596&0.839&0.987\\
				2008&51&0&2&1&256&4.68&3.508&&2015&52&0&3&1&592&0.426&0.927\\
				2009&32&0&3&1&224&4.63&3.844&&2015&53&0&2&1&218&1.459&1.226\\
				2009&33&0&2&1&431&3.386&6.38&&2016&31&0&4&1&106&1.034&0.465\\
				2009&52&0&4&1&355&9.198&13.484&&2016&32&0&2&1&381&1.122&1.250\\
				2010&32&0&2&4&274&1.314&1.168&&2016&33&0&4&2&560&0.729&1.128\\
				2010&33&0&3&2&573&1.365&2.31&&2016&52&0&5&1&566&2.06&0.520\\
				2010&51&0&4&1&248&0.593&1.303&&2016&56&0&4&1&53&0.585&1.370\\
				2010&56&0&5&1&46&2.192&0.647&&2017&21&0&4&3&120&3.634&1.236\\
				2011&52&0&5&1&464&1.284&0.88&&2017&31&0&4&2&106&1.029&1.647\\
				2011&56&0&1&1&49&1.882&1.065&&2017&32&0&1&1&389&1.922&0.502\\
				2012&22&0&5&2&97&0.895&0.73&&2017&33&0&1&1&578&1.834&1.942\\
				2012&32&0&1&2&275&0.517&1.154&&2017&51&0&1&1&331&0.793&0.761\\
				2012&33&0&2&4&572&1.416&2.699&&2017&52&0&4&1&609&0.801&0.538\\
				2012&45&0&1&2&35&7.648&5.01&&2017&53&0&5&1&225&1.639&2.989\\
				2012&51&0&2&2&264&2.309&0.707&&2017&72&0&5&1&63&0.713&1.679\\
				2012&52&0&4&1&499&2.421&1.02&&2018&31&0&5&2&57&2.133&1.103\\
				2012&54&0&1&2&106&1.864&2.355&&2018&32&0&4&2&221&1.748&1.795\\
				2013&31&0&2&1&113&0.574&1.039&&2018&52&0&4&2&429&1.097&0.61\\
				2013&32&0&5&5&292&1.329&1.125&\\
				\hline	
				
			\end{tabular}
		\end{tiny}
	\end{center}
	\raggedright
	\footnotesize{Notes: Models are fitted using the generalised synthetic control method of \cite{xu2017generalized}. NAICS2 reports the two-digit North American Industry Classification System (NAICS) code for the considered industry. MSPE is the Mean Squared Prediction Error when fitting the synthetic versions of the fitted shares to the training data. CAPM reports the MSPE for the CAPM based CARs from Section \ref{sec:simpcar}, whilst Synth reports the MSPE for the generalised synthetic control methodology. V. reports a test for the cointegration of the error matrix with 0 implying rejection. Co. gives the number of cointegrating relationships used in the construction of the unobserved parameter. Tr is the number of firms who exited the DJSI for that two digit NAICS code. Ctrl is the number of control firms used to construct the counterfactual model for de-listed firms. All firms with missing data are eliminated, including some de-listings from the DJSI. All estimations performed using \textit{gsynth} \citep{gsynth}}
\end{table}

Of particular interest to the study of net listing effects are the abnormal returns of periods that involve the announcement date. However to fully evidence any pre-announcement and correction effects we also consider the week immediately before the announcement and the week following the effective date. Start dates are provided from the first day of the treatment period up to, and including the day after the effective date. A full set of abnormal returns over periods of one day, or longer, is included within the appendix. Ahead of the announcement there would be little public information as to which shares are to gain listing. However, the first columns of Table \ref{tab:exit2} reveal a number of positive significant CARs. This applies to periods beginning two weeks prior to the announcement and is most striking for those which end on day 12. Investors who react to these gains by then purchasing the joining stocks would see a significant negative CAR for the period between day 13 and the announcement date. In terms of the extant literature the positive returns are in line with the pre-announcement effect. 

\begin{table}
	\begin{center}
		\caption{Synthetic Control Cumulative Abnormal Returns t-test Summaries}
		\label{tab:exit2}
		\begin{tiny}
			\begin{tabular}{l c c c c c c c c c c c}
				\hline
				From/To & 11 & 12 & 13 & 14 & 15 & ANN & 17 & 18 & 19 & 20 & EFF \\
				\hline
				1&0.353&0.28&0.259&0.364&0.516*&0.598*&0.491&0.4&0.328&0.356&0.572\\
				2&0.399*&0.326&0.305&0.41&0.562*&0.644**&0.537&0.446&0.374&0.402&0.618\\
				3&0.29&0.217&0.196&0.301&0.453&0.535*&0.428&0.337&0.265&0.293&0.509\\
				4&0.194&0.122&0.101&0.206&0.358&0.44&0.333&0.242&0.169&0.197&0.414\\
				5&0.167&0.094&0.074&0.178&0.33&0.412&0.305&0.214&0.142&0.17&0.387\\
				6&0.181&0.108&0.088&0.192&0.345&0.427&0.32&0.228&0.156&0.184&0.401\\
				7&&-0.073&-0.094&0.011&0.163&0.245&0.138&0.047&-0.025&0.003&0.219\\
				8&&&-0.021&0.084&0.236&0.318&0.211&0.12&0.048&0.076&0.292\\
				9&&&&0.105&0.257&0.339*&0.232&0.141&0.068&0.096&0.313\\
				10&&&&&0.152&0.234&0.127&0.036&-0.036&-0.008&0.208\\
				11&&&&&&0.082&-0.025&-0.116&-0.188&-0.16&0.056\\
				12&&&&&&&-0.107&-0.198&-0.27*&-0.242&-0.026\\
				13&&&&&&&&-0.091&-0.163&-0.135&0.081\\
				14&&&&&&&&&-0.072&-0.044&0.172\\
				15&&&&&&&&&&0.028&0.245\\
				16&&&&&&&&&&&0.217\\
				& ANN & 17 & 18 & 19 & 20 & EFF & 22 & 23 & 26 & 31 & 36 \\ 
				1&0.572&0.608&0.715*&0.89*&0.742&0.713&0.556&0.534&0.512&0.33&0.441\\
				2&0.618&0.654&0.762*&0.936**&0.789&0.76&0.602&0.58&0.558&0.376&0.487\\
				3&0.509&0.545&0.653&0.827*&0.68&0.65&0.493&0.471&0.449&0.267&0.378\\
				4&0.414&0.45&0.557&0.732*&0.584&0.555&0.397&0.376&0.354&0.172&0.283\\
				5&0.387&0.422&0.53&0.704&0.557&0.528&0.37&0.348&0.326&0.144&0.256\\
				6&0.401&0.436&0.544&0.719&0.571&0.542&0.384&0.363&0.34&0.159&0.270\\
				7&0.219&0.255&0.363&0.537&0.39&0.361&0.203&0.181&0.159&-0.023&0.088\\
				8&0.292&0.328&0.436&0.61&0.463&0.434&0.276&0.254&0.232&0.05&0.161\\
				9&0.313&0.349&0.456&0.631&0.483&0.454&0.296&0.275&0.253&0.071&0.182\\
				10&0.208&0.244&0.352&0.526&0.379&0.35&0.192&0.17&0.148&-0.034&0.077\\
				11&0.056&0.092&0.2&0.374&0.227&0.197&0.04&0.018&-0.004&-0.186&-0.075\\
				12&-0.026&0.01&0.117&0.292&0.144&0.115&-0.042&-0.064&-0.086&-0.268&-0.157\\
				13&0.081&0.117&0.224&0.399&0.251&0.222&0.064&0.043&0.021&-0.161&-0.050\\
				14&0.172&0.208&0.316&0.49&0.343&0.313&0.156&0.134&0.112&-0.07&0.041\\
				15&0.245&0.28&0.388&0.562*&0.415&0.386&0.228&0.206&0.184&0.002&0.114\\
				ANN&0.217&0.252&0.36&0.534*&0.387&0.358&0.2&0.178&0.156&-0.026&0.086\\
				17&&0.036&0.143&0.318&0.17&0.141&-0.017&-0.038&-0.06&-0.242&-0.131\\
				18&&&0.108&0.282&0.135&0.106&-0.052&-0.074&-0.096&-0.278&-0.167\\
				19&&&&0.175&0.027&-0.002&-0.16&-0.181&-0.204&-0.385&-0.274\\
				20&&&&&-0.148&-0.177&-0.334&-0.356&-0.378&-0.56*&-0.449\\
				EFF&&&&&&-0.029&-0.187&-0.208&-0.231&-0.412&-0.301\\
				22&&&&&&&-0.158&-0.179&-0.201&-0.383&-0.272\\
			\end{tabular}
		\end{tiny}
	\end{center}
\raggedright
\footnotesize{Notes: Average cumulative abnormal returns are reported for the period starting on the row label and ending according to the column label. These are averaged over time and industry. A t-test across the time-industry space results in a report of their difference from zero. Significant returns are denoted by * - 5\%, ** - 1\% and *** - 0.1\%. A full set of results are reported in the supplementary material.}
\end{table}

Evidence is provided that much of the uptick from DJSI listing occurs prior to the change date, a result which appears in \cite{oberndorfer2013does}. There is also evidence in the synthetic control cumulative abnormal returns of the correction that \cite{oberndorfer2013does} argues takes place after the announcement date. Such reversion effects manifest as negative abnormal returns in the windows starting after the announcement date and ending in the trading days immediately after the effective date. Three weeks post effective date none of the reduced set of periods show significant cumulative abnormal returns. Herein we see the correction effect mentioned in the literature. But, as can be seen in Table \ref{tab:avg2} this is not as strong as it was implied to be by the CAPM in Table \ref{tab:cars}. For all time frames starting before the announcement date there are positive cumulative abnormal returns but their magnitude diminishes and their significance is lost post announcement. 

Using the same reduced set of from and to dates we may appraise the magnitude of the abnormal returns on holding shares in firms which de-list from the DJSI in the given years. As with the listings there is a clear pre-announcement effect with positive CARs are seen for holding periods which end in the days prior to the announcement. Compared to the listing effects these pre-announcement values have greater significance. Many ranges which end on the announcement date have significance for de-listing, but this was not seen in the listing case. Once the information about which firms have not met the criteria to retain their place on the DJSI becomes public there are far fewer statistically significant differences between average CARs and zero compared to that for the firms announced as joining. Strong negative correction effects appear throughout the lower right of Table \ref{tab:exit2} but none of these have any statistical significance attached.

\begin{table}
	\begin{center}
		\caption{Synthetic Control Cumulative Abnormal Returns t-test Summaries: De-listing}
		\label{tab:avg2}
		\begin{tiny}
			\begin{tabular}{l c c c c c c c c c c c}
				\hline
				From/To & 6&7&8&9&10&11&12&13&14&15 & ANN \\
				\hline
				1 & -0.345&-0.157&-0.02&0.24&0.455&0.469&0.417&0.61&0.566&0.495&0.605\\
				2&-0.194&-0.006&0.131&0.391&0.606&0.62&0.568&0.761&0.717&0.646&0.756\\
				3&-0.107&0.081&0.218&0.478&0.693&0.707&0.654&0.848*&0.804*&0.733&0.843*\\
				4&-0.074&0.114&0.251&0.512&0.726*&0.741*&0.688&0.882*&0.837*&0.766&0.877*\\
				5&-0.16&0.028&0.165&0.425&0.64&0.654&0.602&0.795*&0.751&0.68&0.790\\
				6&-0.068&0.12&0.256&0.517&0.731*&0.746*&0.693*&0.887*&0.842*&0.771&0.882*\\
				7&&0.188&0.325&0.585*&0.8*&0.814**&0.762*&0.955**&0.911**&0.84*&0.950**\\
				8&&&0.137&0.397&0.612*&0.626*&0.574&0.767*&0.723*&0.652&0.762*\\
				9&&&&0.261&0.475&0.489&0.437&0.63&0.586&0.515&0.626\\
				10&&&&&0.214&0.229&0.176&0.37&0.325&0.254&0.365\\
				11&&&&&&0.015&-0.038&0.155&0.111&0.04&0.151\\
				12&&&&&&&-0.053&0.141&0.096&0.025&0.136\\
				13&&&&&&&&0.194&0.149&0.078&0.189\\
				14&&&&&&&&&-0.044&-0.115&-0.005\\
				15&&&&&&&&&&-0.071&0.040\\
				ANN&&&&&&&&&&&0.111\\
				& ANN & 17 & 18 & 19 & 20 & EFF & 22 & 23 & 26 & 31 & 36\\ 
				1 & 0.605&0.709&0.741&0.799&0.531&0.453&0.467&0.173&0.217&0.308&0.214\\
				2 & 0.756&0.859*&0.892*&0.95*&0.682&0.604&0.618&0.324&0.367&0.459&0.364\\
				3 & 0.843*&0.946*&0.979*&1.036**&0.769&0.691&0.704&0.411&0.454&0.545&0.451\\
				4 & 0.877*&0.98*&1.012**&1.07**&0.802&0.724&0.738&0.445&0.488&0.579&0.485\\
				5 & 0.790&0.893*&0.926*&0.984*&0.716&0.638&0.652&0.358&0.401&0.493&0.398\\
				6 & 0.882*&0.985**&1.017**&1.075**&0.808&0.729&0.743&0.45&0.493&0.584&0.490\\
				7 & 0.950**&1.053**&1.086**&1.144**&0.876&0.798&0.812&0.518&0.561&0.653&0.558\\
				8 & 0.762*&0.866*&0.898**&0.956**&0.688&0.61&0.624&0.33&0.373&0.465&0.371\\
				9 & 0.626&0.729*&0.761*&0.819*&0.551&0.473&0.487&0.193&0.237&0.328&0.234\\
				10 & 0.365&0.468&0.501&0.558&0.291&0.212&0.226&-0.067&-0.024&0.067&-0.027\\
				11 & 0.151&0.254&0.286&0.344&0.076&-0.002&0.012&-0.282&-0.238&-0.147&-0.241\\
				12 & 0.136&0.239&0.272&0.329&0.062&-0.017&-0.003&-0.296&-0.253&-0.162&-0.256\\
				13 & 0.189&0.292&0.324&0.382&0.114&0.036&0.05&-0.244&-0.2&-0.109&-0.203\\
				14 & -0.005&0.098&0.131&0.188&-0.079&-0.158&-0.144&-0.437&-0.394&-0.303&-0.397\\
				15 & 0.040&0.143&0.175&0.233&-0.035&-0.113&-0.099&-0.393&-0.349&-0.258&-0.352\\
				ANN & 0.111&0.214&0.246&0.304*&0.036&-0.042&-0.028&-0.322&-0.278&-0.187&-0.281\\
				17&&0.103&0.136&0.193&-0.074&-0.153&-0.139&-0.432&-0.389&-0.298&-0.392\\
				18&&&0.032&0.09&-0.178&-0.256&-0.242&-0.535&-0.492&-0.401&-0.495\\
				19&&&&0.058&-0.21&-0.288&-0.274&-0.568&-0.524&-0.433&-0.527\\
				20&&&&&-0.268&-0.346&-0.332&-0.626&-0.582&-0.491&-0.585\\
				EFF&&&&&&-0.078&-0.064&-0.358&-0.314&-0.223&-0.317\\
				22&&&&&&&0.014&-0.28&-0.236&-0.145&-0.239\\
			\end{tabular}
		\end{tiny}
	\end{center}
	\raggedright
	\footnotesize{Notes: Average cumulative abnormal returns are reported for the period starting on the row label and ending according to the column label. These are averaged over time and industry. A t-test across the time-industry space results in a report of their difference from zero. Significant returns are denoted by * - 5\%, ** - 1\% and *** - 0.1\%. A full set of results are reported in the supplementary material.}
\end{table}

\subsection{Comparison}
\label{sec:compare}

Three approaches to identifying the listing effects of joining the DJSI have been presented in this paper and we have seen variations in the predictions made. Two-sample tests revealed few significant effects pre-listing with only a small region of positive CARs identified a week before the announcement. Holdings from this time offered significant returns when held to day 19 also. When considering the base sample these joining effects lose significance. Only those positive CARs on periods starting on the announcement date, and ending on the subsequent days, are significant in both the full and base samples. None of these positive listing effects are noted in the OLS modelling, subsumed in the control for leverage and profitability. Given the large size requirement to join the DJSI such results may seem unsurprising, but they are premised on the idea that out-of-sample prediction from simple asset pricing models in the correct way to calculate expected returns.

In the OLS regressions there were no significant impacts of listing for any of the considered periods. Identified CARs were absorbed by the strength of the role of other firm characteristics that are linked to returns; size leverage and profitability were all shown to have significance in Table \ref{tab:carlm}. Critically gaining DJSI status was not. Such significance shows the importance of the controls when comparing two samples and reminds of the need to take care when using differences in sample means as measures of impact. 

Utilising the generalised synthetic control approach means there is only one variable being considered, the returns of the share for which the synthetic control is being generated. What is important for generating the counterfactual is not the levels of financial performance in the control set, but the way their share price contributes to a portfolio which matches the behaviour of the firm to be listed. Like the CAPM we have only to consider the firms at an individual level, but the presence of the controls is allowing the performance of others to affect CARs. Across a single year there will be few changes in a firms financial performance, meaning that the relationships between shares would not be expected to change by much between the control and treatment periods. Hence the CARs encapsulate many of the benefits that come from understanding the relative properties of the control firms without imposing restrictions on the control set.

When considering the de-listing effects a further comparison may be drawn between these results and those obtained from the CAPM CARs in Table \ref{tab:carsx}. Pre-announcement effects are much larger in magnitude and significant for a far greater range of dates than they were in the full sample comparison. When restricting to the base sample there are more similarities; such is natural as the algorithm creates a weighted portfolio to match control period returns and would therefore select firms with closer characteristics to the de-listing firm. However, the magnitude of the generalised synthetic control differentials is slightly smaller and fades rapidly post announcement; the CAPM CARs from the base sample did not. To the study of firms that leave the DJSI the generalised synthetic control approach thus offers a mediator between effects that are overstated in the base sample and understated for the full sample, doing so whilst drawing on the maximum possible information set.

Within the results there is thus support for the twin hypotheses derived from the literature. Firstly the benefit of higher expected profitability causes a rise in the price of the share over and above any effects happening to the control shares. Secondly there is a correction as the lower risk associated with CSR activities takes over; such is consistent with the standard assumptions on the risk/return relationship. Our confirmation from this approach lends weight to the theories and results of \cite{oberndorfer2013does} and others.

Our generalised synthetic control approach picks up a strong positive significant return on the listed firms if held from the start of the treatment period to the week ahead of the announcement, and indeed on the same shares held to the effective date. In this way the pre-announcement effect is more notable than it is in the CAPM based bivariate analyses. By contrast the large gains that were suggested between announcement and effective dates are not significant and much smaller in magnitude. As with the pre-announcement effect, the correction effect is much stronger in the synthetic control. Buying the listed firms the day after listing consistently produces negative CARs; shorting these shares augurs potential profit for investors. This paper thus evidences more effects of gaining DJSI listing than past approaches. In so doing it reinforces the need for investors to research carefully when considering buying firms which achieve, or are likely to achieve, the recognition of good CSR practice afforded by DJSI listing.

\section{Year Effects}

Heterogeneity between years is readily controlled in this analysis as the process of abnormal return generation treats each year independently. Using the data collected tests can be performed to identify whether the average CAR for a given holding period is identical across two years. By constructing matched pairs over a week of start dates and a week of end dates subsamples are created that may then be considered. In what follows the average difference is reported together with a significance based upon the t-test for equality between the paired samples. First the tests are performed for the firms gaining listing to the DJSI and second for those being de-listed. For this element of the paper effects are only studied for the generalised synthetic control returns. Theory may suggest that investor reaction to these two opposite events would be in opposite directions. 

Tables \ref{tab:yrlist} and \ref{tab:yrdelist} present t-tests for equality of CARs on each holding start and end date between year pairs. These are paired t-tests since the aim is to establish equality for a given period, for example from announcement day to effective day. Each panel therefore considers a range of from and to dates to create the samples for testing. Panel (a) looks at stocks being purchased in the first week of the treatment period, Week 1 (days 1 to 5 inclusive), and being held to the week before the announcement, Week 3 (days 11 to 14), inclusive. Panel (a) thus covers the longest period of the samples considered. It was seen that there were strong positive returns when averaging across the years. Panel (b) focuses in on the days prior to listing considering only those holding periods starting, and ending, between days 11 and 14 inclusive. Panel (c) includes the day before the announcement and runs through to the day after the effective day. In this way all impacts concerned with the publication of information are captured. Finally, panel (d) looks at the period immediately subsequent to this to capture any correction effects that may occur. Identical periods are used here for both listing and de-listing.

\begin{sidewaystable}
	\begin{center}
		\caption{New way of doing things}
		\label{tab:yrlist}
		\begin{tiny}
			\begin{tabular}{l c c c c c c c c c c c c c}
				\hline
				from/to & 2006 & 2007 & 2008 & 2009 & 2010 & 2011 & 2012 & 2013 & 2014 & 2015 & 2016 & 2017 & 2018 \\
				\hline
				2005 & 0.491***&-2.196***&-0.385&0.642***&-2.271***&-1.34***&-0.528***&-0.004&-0.518*&0.208&-0.233**&1.362***&-0.234\\
				2006&&-2.687***&-0.876*&0.151&-2.761***&-1.831***&-1.019***&-0.494***&-1.009***&-0.283&-0.723***&0.871***&-0.725***\\
				2007&&&1.811***&2.838***&-0.074&0.856***&1.668***&2.193***&1.678***&2.404***&1.964***&3.558***&1.962***\\
				2008&&&&1.027*&-1.885**&-0.955*&-0.143&0.382&-0.133&0.593&0.153&1.748***&0.151\\
				2009&&&&&-2.913***&-1.982***&-1.17***&-0.646***&-1.16***&-0.434***&-0.875***&0.72***&-0.876***\\
				2010&&&&&&0.93***&1.742***&2.267***&1.753***&2.478***&2.038***&3.633***&2.037***\\
				2011&&&&&&&0.812***&1.337***&0.822**&1.548***&1.108***&2.703***&1.106***\\
				2012&&&&&&&&0.525***&0.01&0.736***&0.296***&1.89***&0.294*\\
				2013&&&&&&&&&-0.515&0.211&-0.229*&1.366***&-0.230**\\
				2014&&&&&&&&&&0.726&0.285&1.88***&0.284\\
				2015&&&&&&&&&&&-0.44**&1.155***&-0.442**\\
				2016&&&&&&&&&&&&1.595***&-0.001\\
				2017&&&&&&&&&&&&&-1.596***\\
				\hline
				
				2005&0.201**&-0.706*&1.451*&-0.262&-1.085***&-0.366*&0.048&-0.588*&0.03&-0.085&-0.149&0.209&-0.74**\\
				2006&&-0.505&1.652**&-0.061&-0.884***&-0.165&0.249*&-0.387*&0.231&0.116&0.052&0.41**&-0.539**\\
				2007&&&2.157*&0.444&-0.379&0.34&0.754*&0.118&0.736*&0.621*&0.558&0.915**&-0.034\\
				2008&&&&-1.714*&-2.536**&-1.817*&-1.403*&-2.039*&-1.422*&-1.537*&-1.6**&-1.242*&-2.191**\\
				2009&&&&&-0.822***&-0.103&0.31&-0.326**&0.292*&0.177*&0.114&0.472**&-0.477***\\
				2010&&&&&&0.719***&1.133***&0.497**&1.115***&1***&0.936***&1.294***&0.345*\\
				2011&&&&&&&0.414&-0.223&0.395*&0.28&0.217&0.575***&-0.374*\\
				2012&&&&&&&&-0.636**&-0.018&-0.133&-0.196**&0.161&-0.787**\\
				2013&&&&&&&&&0.618***&0.503***&0.44*&0.798**&-0.151\\
				2014&&&&&&&&&&-0.115*&-0.178&0.18&-0.769**\\
				2015&&&&&&&&&&&-0.063&0.295*&-0.654**\\
				2016&&&&&&&&&&&&0.358*&-0.591*\\
				2017&&&&&&&&&&&&&-0.949***\\
				\hline
				2005&-0.085&-0.148&-1.896***&-0.156&-0.155&-0.028&0.418**&0.173&-0.565***&-0.579***&-0.17&0.309*&0.755***\\
				2006&&-0.063&-1.811***&-0.07&-0.07&0.057&0.503***&0.259&-0.48***&-0.494***&-0.084&0.395***&0.84***\\
				2007&&&-1.749***&-0.008&-0.008&0.119&0.566***&0.321**&-0.417***&-0.431**&-0.022&0.457***&0.903***\\
				2008&&&&1.741***&1.741***&1.868***&2.314***&2.07***&1.331***&1.317***&1.727***&2.206***&2.651***\\
				2009&&&&&&0.127&0.573***&0.329&-0.41***&-0.424***&-0.014&0.465**&0.91***\\
				2010&&&&&&0.127&0.573***&0.329*&-0.41***&-0.424**&-0.014&0.465***&0.91***\\
				2011&&&&&&&0.446*&0.202&-0.537**&-0.551*&-0.141&0.338*&0.783***\\
				2012&&&&&&&&-0.244&-0.983***&-0.997***&-0.587***&-0.108&0.337**\\
				2013&&&&&&&&&-0.739***&-0.752**&-0.343*&0.136&0.582**\\
				2014&&&&&&&&&&-0.014&0.396***&0.875***&1.32***\\
				2015&&&&&&&&&&&0.409**&0.889***&1.334***\\
				2016&&&&&&&&&&&&0.479***&0.925***\\
				2017&&&&&&&&&&&&&0.446***\\
				\hline
				2005&-0.611***&-0.65***&-1.906***&-0.729***&-1.011***&-0.282*&0.166&0.867***&-0.32*&-1.967***&-0.845***&-0.353**&1.462***\\
				2006&0&-0.039&-1.295***&-0.118&-0.4***&0.329*&0.777***&1.478***&0.291&-1.356***&-0.234**&0.257**&2.072***\\
				2007&&&-1.256***&-0.079&-0.361***&0.368***&0.816***&1.517***&0.33*&-1.317***&-0.195**&0.297**&2.112***\\
				2008&&&&1.177***&0.895***&1.624***&2.072***&2.773***&1.586***&-0.061&1.061***&1.553***&3.368***\\
				2009&&&&&-0.282**&0.447***&0.895***&1.596***&0.409**&-1.238***&-0.116&0.376**&2.191***\\
				2010&&&&&&0.729***&1.177***&1.878***&0.691***&-0.956***&0.166*&0.658***&2.473***\\
				2011&&&&&&&0.448**&1.149***&-0.038&-1.685***&-0.563***&-0.071&1.744***\\
				2012&&&&&&&&0.701***&-0.486**&-2.133***&-1.011***&-0.519***&1.296***\\
				2013&&&&&&&&&-1.187***&-2.834***&-1.712***&-1.221***&0.594***\\
				2014&&&&&&&&&&-1.647***&-0.525***&-0.034&1.782***\\
				2015&&&&&&&&&&&1.122***&1.613***&3.429***\\
				2016&&&&&&&&&&&&0.492***&2.307***\\
				2017&&&&&&&&&&&&&1.815***\\
				\hline
			\end{tabular}
		\end{tiny}
	\end{center}
\end{sidewaystable}

For firms joining the DJSI panel (a) of Table \ref{tab:yrlist} shows a larger CAR for 2007 as the market learns more about what listing means.  However, in 2008 there is no significance and 2009 is regularly smaller than later years. Only further into the post crisis period are the average CARs larger. Positive values in the 2017 and 2018 columns inform that recent listings have not had such large effects as the years described by the rows. Such patterns extend into the consideration of just the pre-announcement week (panel (b)), although here 2007 is less significant and the difference between 2008 and other years is much larger. Panel (c) shows further dominance by 2008 but moving toward the bottom right it is apparent that the row years often have smaller CARs than the column years; the size of the listing CAR is growing over time. Similar ``u-shaped'' relations are also spotted in the post-listing period of panel (d). Here again the financial crisis has strong positive values to it's t-test difference with each local firm. From around 2010 all years have smaller average CARs than the subsequent year; this may mean the market is learning but heterogeneity across firms limits the depth to which such differentials may be read into.

\begin{sidewaystable}
	\begin{center}
		\caption{Two-sample tests for CAR equality between years: Delisting}
		\label{tab:yrdelist}
		\begin{tiny}
			\begin{tabular}{l c c c c c c c c c c c c }
				\hline
				from/to & 2007 & 2008 & 2009 & 2010 & 2011 & 2012 & 2013 & 2014 & 2015 & 2016 & 2017 & 2018 \\
				\hline
				\multicolumn{13}{l}{Panel(a): Week 1 to Week 3}\\
				2006 &0.897**&2.019***&0.708***&0.271**&1.139***&0.263*&0.891***&1.647***&1.049***&1.223***&0.615***&1.286***"\\
				2007&&1.123***&-0.189&-0.625*&0.242&-0.634&-0.005&0.750**&0.153&0.327&-0.281&0.389\\
				2008&&&-1.311***&-1.748***&-0.881***&-1.756***&-1.128***&-0.372&-0.97***&-0.796***&-1.404***&-0.734***\\
				2009&&&&-0.437***&0.431**&-0.445***&0.183&0.939**&0.341&0.515**&-0.093&0.578***\\
				2010&&&&&0.867***&-0.008&0.62***&1.376***&0.778***&0.952***&0.344**&1.014***\\
				2011&&&&&&-0.876***&-0.247***&0.508*&-0.089&0.085&-0.523**&0.147*\\
				2012&&&&&&&0.628***&1.384***&0.787***&0.96***&0.352**&1.023***\\
				2013&&&&&&&&0.756**&0.158&0.332*&-0.276*&0.394***\\
				2014&&&&&&&&&-0.597*&-0.424&-1.032**&-0.361\\
				2015&&&&&&&&&&0.174&-0.434&0.236\\
				2016&&&&&&&&&&&-0.608***&0.062\\
				2017&&&&&&&&&&&&0.67***\\
				\hline
				\multicolumn{13}{l}{Panel(b): Week 3 to Week 3}\\
				2006 & 0.595**&1.945**&0.8*&0.284*&0.024&0.129&0.194*&0.398**&0.772***&-1.018**&-0.253**&0.175\\
				2007&&1.35*&0.205&-0.311**&-0.571**&-0.465**&-0.401&-0.197*&0.178&-1.612***&-0.848**&-0.420\\
				2008&&&-1.145*&-1.661**&-1.921**&-1.815**&-1.751*&-1.547*&-1.172*&-2.962**&-2.198**&-1.77*\\
				2009&&&&-0.516&-0.776&-0.671&-0.606&-0.402&-0.028&-1.818**&-1.053*&-0.625\\
				2010&&&&&-0.26*&-0.155*&-0.09&0.114&0.488*&-1.302**&-0.537**&-0.109\\
				2011&&&&&&0.105&0.17&0.373*&0.748*&-1.042**&-0.277&0.151\\
				2012&&&&&&&0.065&0.268*&0.643**&-1.147**&-0.382*&0.046\\
				2013&&&&&&&&0.204&0.578***&-1.212**&-0.447**&-0.019\\
				2014&&&&&&&&&0.375*&-1.415**&-0.65**&-0.222\\
				2015&&&&&&&&&&-1.790***&-1.025***&-0.597**\\
				2016&&&&&&&&&&&0.765*&1.193**\\
				2017&&&&&&&&&&&&0.428**\\
				\hline
				\multicolumn{13}{l}{Panel(c): Week 4 to Week 4}\\
				2006&0.512***&1.338***&1.744***&1.155***&1.218***&0.728***&1.195***&0.658***&0.240**&1.047***&0.803***&0.603**\\
				2007&&0.826**&1.232***&0.643***&0.707***&0.216*&0.684***&0.146&-0.271**&0.535***&0.291***&0.091\\
				2008&&&0.406&-0.183&-0.119&-0.61*&-0.142&-0.68*&-1.097**&-0.291&-0.535&-0.735*\\
				2009&&&&-0.589***&-0.526**&-1.016***&-0.549***&-1.086***&-1.504***&-0.697***&-0.941***&-1.141***\\
				2010&&&&&0.064&-0.427***&0.04&-0.497***&-0.914***&-0.108&-0.352***&-0.552***\\
				2011&&&&&&-0.49**&-0.023&-0.56***&-0.978***&-0.172&-0.416*&-0.615**\\
				2012&&&&&&&0.467***&-0.07&-0.488**&0.319**&0.075&-0.125\\
				2013&&&&&&&&-0.537***&-0.955***&-0.149**&-0.393***&-0.592***\\
				2014&&&&&&&&&-0.418*&0.389**&0.145&-0.055\\
				2015&&&&&&&&&&0.806***&0.562***&0.363\\
				2016&&0&0&0&0&0&0&0&0&0&-0.244**&-0.444***\\
				2017&&0&0&0&0&0&0&0&0&0&0&-0.200\\
				\hline	
				\multicolumn{13}{l}{Panel(d): Week 4 to Week 5}\\
				2006& 1.784***&-0.395&2.242***&1.965***&3.959***&2.376***&3.389***&1.808***&1.401***&2.033***&1.916***&2.01***\\
				2008&&-2.18***&0.457***&0.18&2.175***&0.591***&1.605***&0.023&-0.384***&0.249*&0.131*&0.226*\\
				2009&&&2.637***&2.36***&4.354***&2.771***&3.785***&2.203***&1.796***&2.428***&2.311***&2.406***\\
				2010&&&&-0.277***&1.717***&0.134&1.148***&-0.434***&-0.841***&-0.209&-0.326**&-0.231*\\
				2011&&&&&1.994***&0.411***&1.425***&-0.157&-0.564***&0.068&-0.049&0.046\\
				2012&&&&&&-1.583***&-0.57***&-2.152***&-2.558***&-1.926***&-2.043***&-1.949***\\
				2013&&&&&&&1.014***&-0.568***&-0.975***&-0.343***&-0.46***&-0.365***\\
				2014&&&&&&&&-1.582***&-1.988***&-1.356***&-1.474***&-1.379***\\
				2015&&&&&&&&&-0.407***&0.225*&0.108&0.203**\\
				2016&&&&&&&&&&0.632***&0.515***&0.61***\\
				2017&&&&&&&&&&&-0.117&-0.022\\
				2018&&&&&&&&&&&&0.095\\
			\end{tabular}
		\end{tiny}
	\end{center}
	\raggedright
	\footnotesize{Notes: CARs . To aid identification weeks are labelled such that Week 1 includes days 1 to 5 inclusive, Week 2 covers days 6 to 10 inclusive and Week 3 covers days 11 to 14 inclusive. Week 4 covers days 15 to 22 inclusive, whilst Week 5 is used to get at the correction effect in days 22 to 27 inclusive.} 
\end{sidewaystable}

An immediate observation is that again there is strong significance in all of the comparisons. 2006 had the largest CARs in all periods, this may be suggestive of learning about the index and the impact on stock prices for a firm being removed from the list. Holdings from the first week of the treatment period (Week 1) to the days before the announcement (Week 3) were notably smaller in 2008 and were then significantly smaller than in all subsequent years. Panel (a) of Table \ref{tab:yrdelist} shows that only 2014 has similarly small CARs as 2008, both being significantly lower than other years. Large positive differences between 2006 and other years are likely to be driven by the declining state of the economy post crisis and the fact that it never fully recovered. There may also be expected to be a realisation that consumers are increasingly likely to punish firms who miss their CSR targets. [**** ADD REFERENCE HERE]

For those periods which include the announcement interest is naturally drawn to the span between announcement and effective date. Panel (c) reports on the period between the announcement and the de-listing revealing very mixed differences in the CARs. Whilst 2009 has consistently lower CARs than more recent years there is little consistency in the direction of the changes between 2012 and 2018. Such changes are in alignment with investors learning about the behaviour of stocks after de-listing and hence gaining a deeper understanding of the implications of a de-listing announcement. Looking at panel (d) it may be seen that more recently the CARs have been larger than they were through the crisis and its' aftermath. Recalling that this panel is studying the correction effect, with negative CARs, this test is showing that the magnitude of the mean reversion is larger in recent years. 

A comparison between the exit results and those for the entering firms informs that the reduction in CARs for the pre-announcement period is broadly common on both listing and de-listing. 2010 and 2011 have bigger CARs on average than are observed from 2014 onwards.Theory suggests that the market would be moving in opposite directions for listing and de-listing since the implications on the CSR signal are in opposite directions. In the early years that is evidenced by the t-tests of Tables \ref{tab:yrdelist}

\begin{figure}
	\begin{center}
		\caption{Comparison of listing and de-listing effects}
		\label{fig:yearcomp}
		\begin{tabular}{c c}
			\includegraphics[width=6cm,height=4cm]{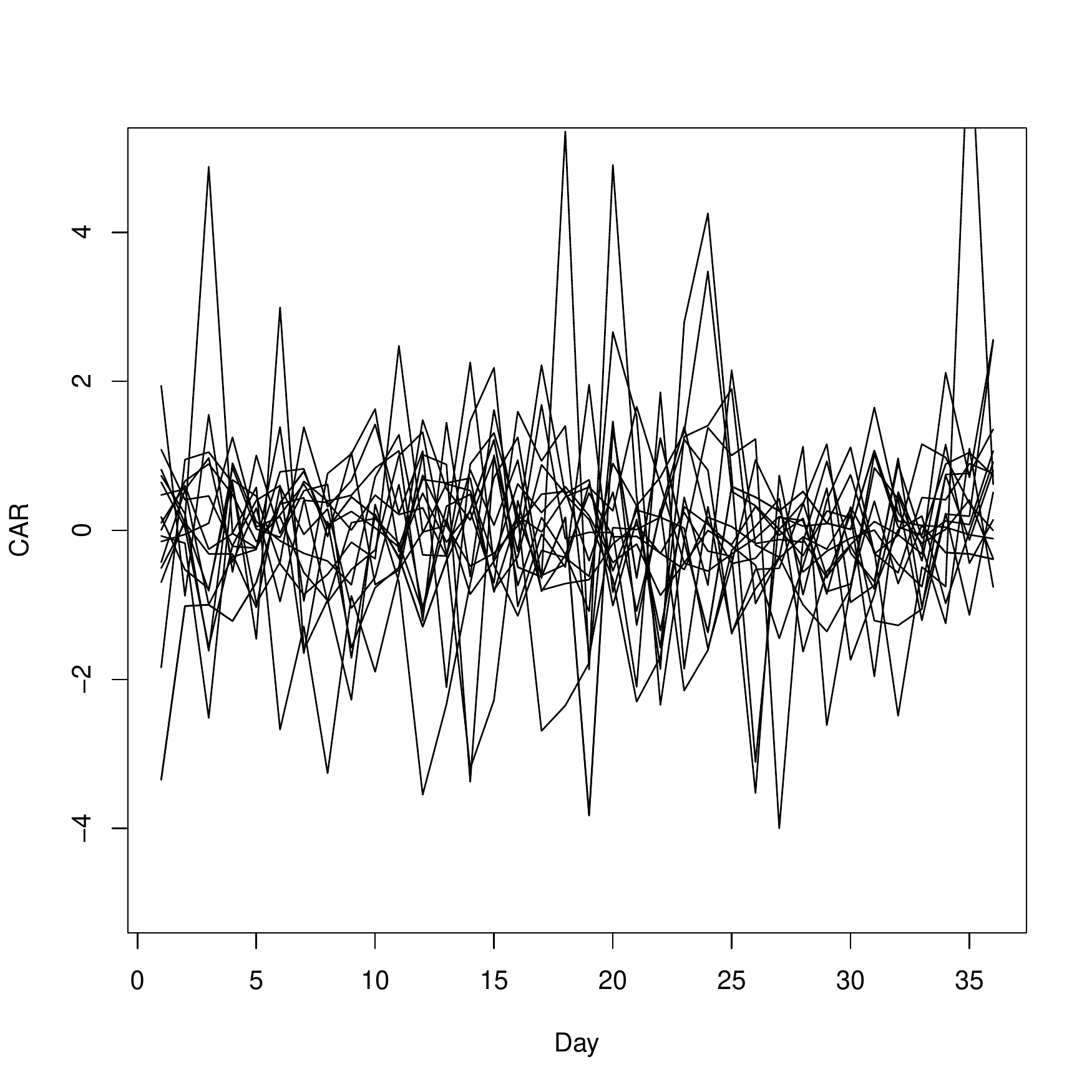} &
			\includegraphics[width=6cm,height=4cm]{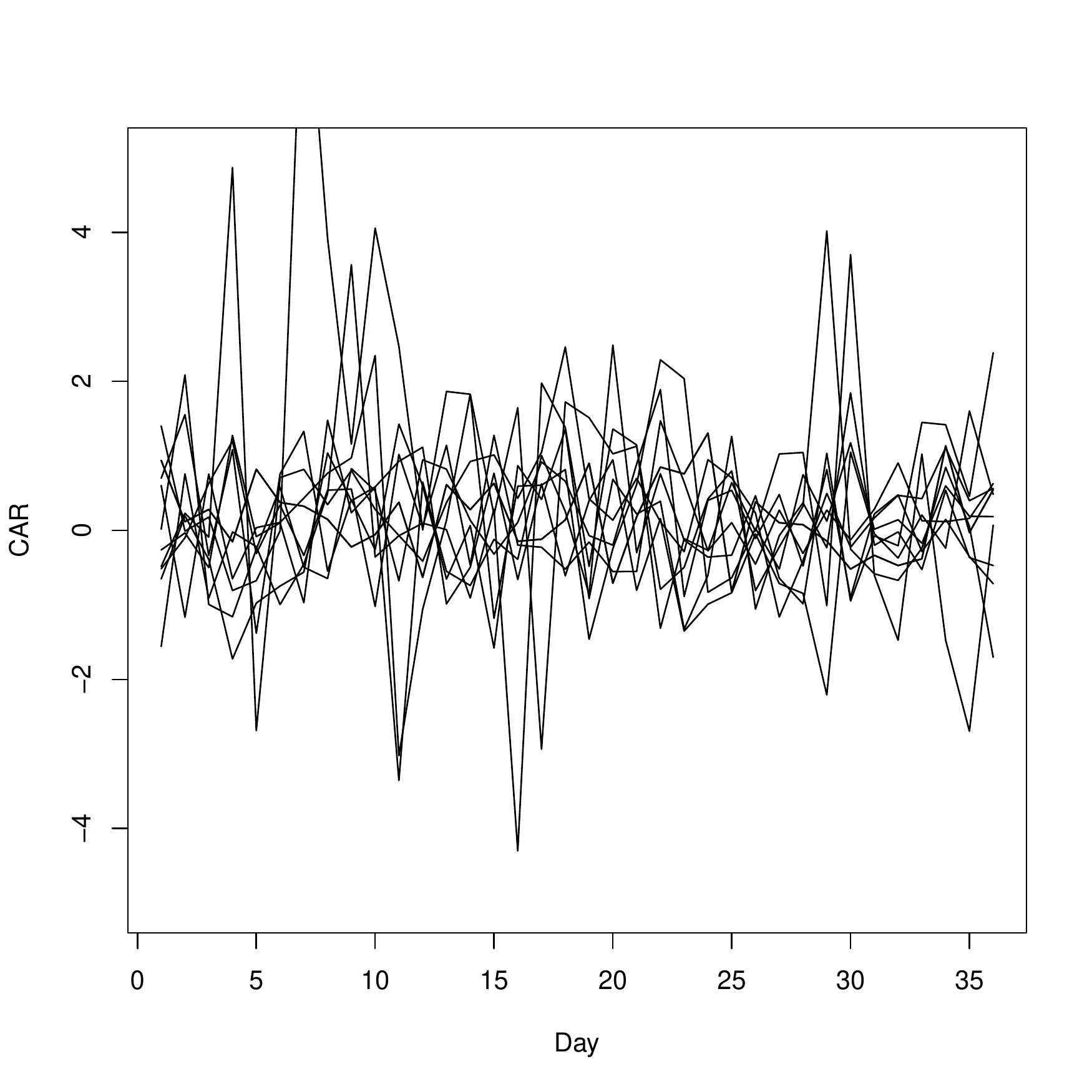} \\
			(a) Listed 2006 & (b) Delisted 2006\\
			\includegraphics[width=6cm,height=4cm]{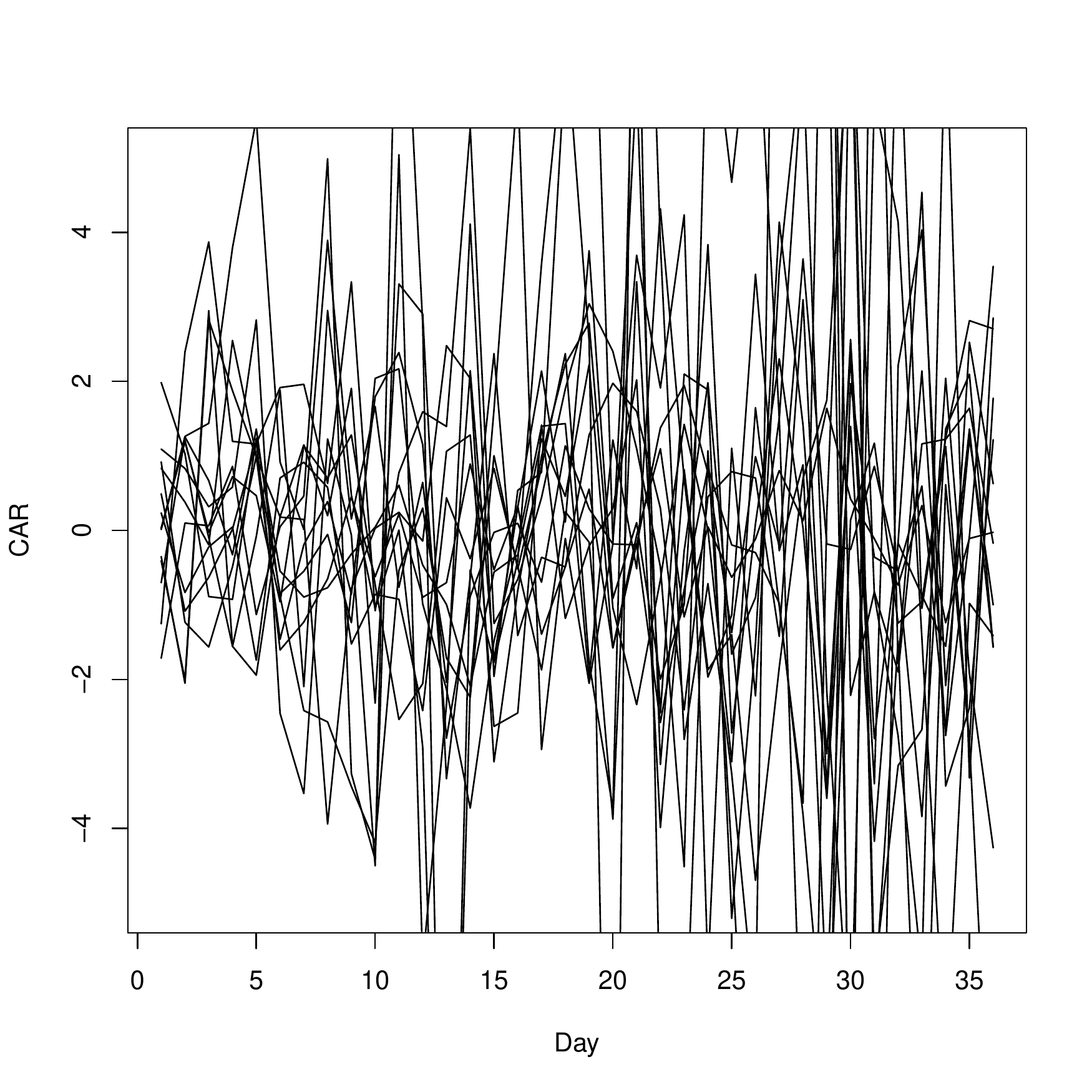} &
			\includegraphics[width=6cm,height=4cm]{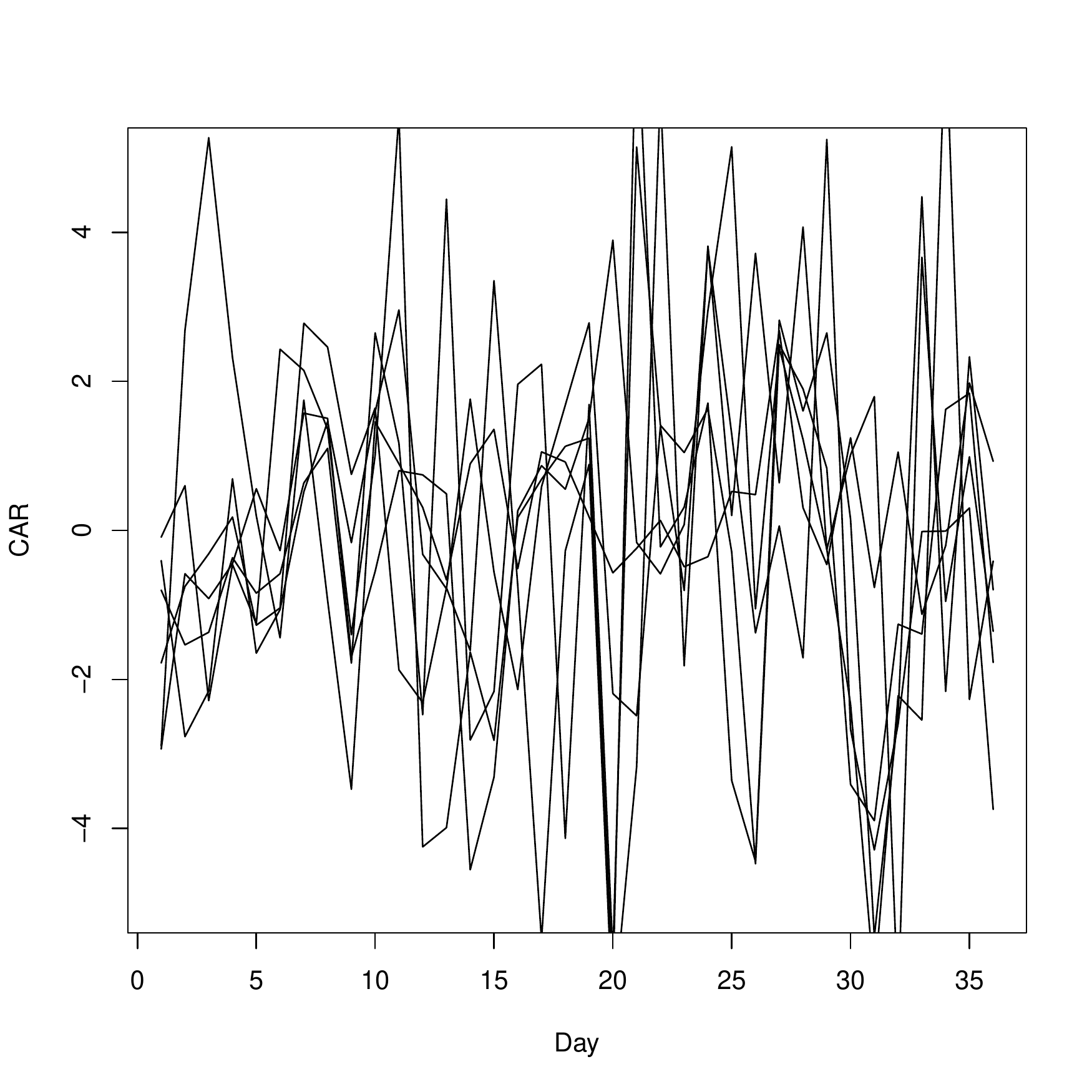} \\
			(c) Listed 2008 & (d) Delisted 2008\\
n			\includegraphics[width=6cm,height=4cm]{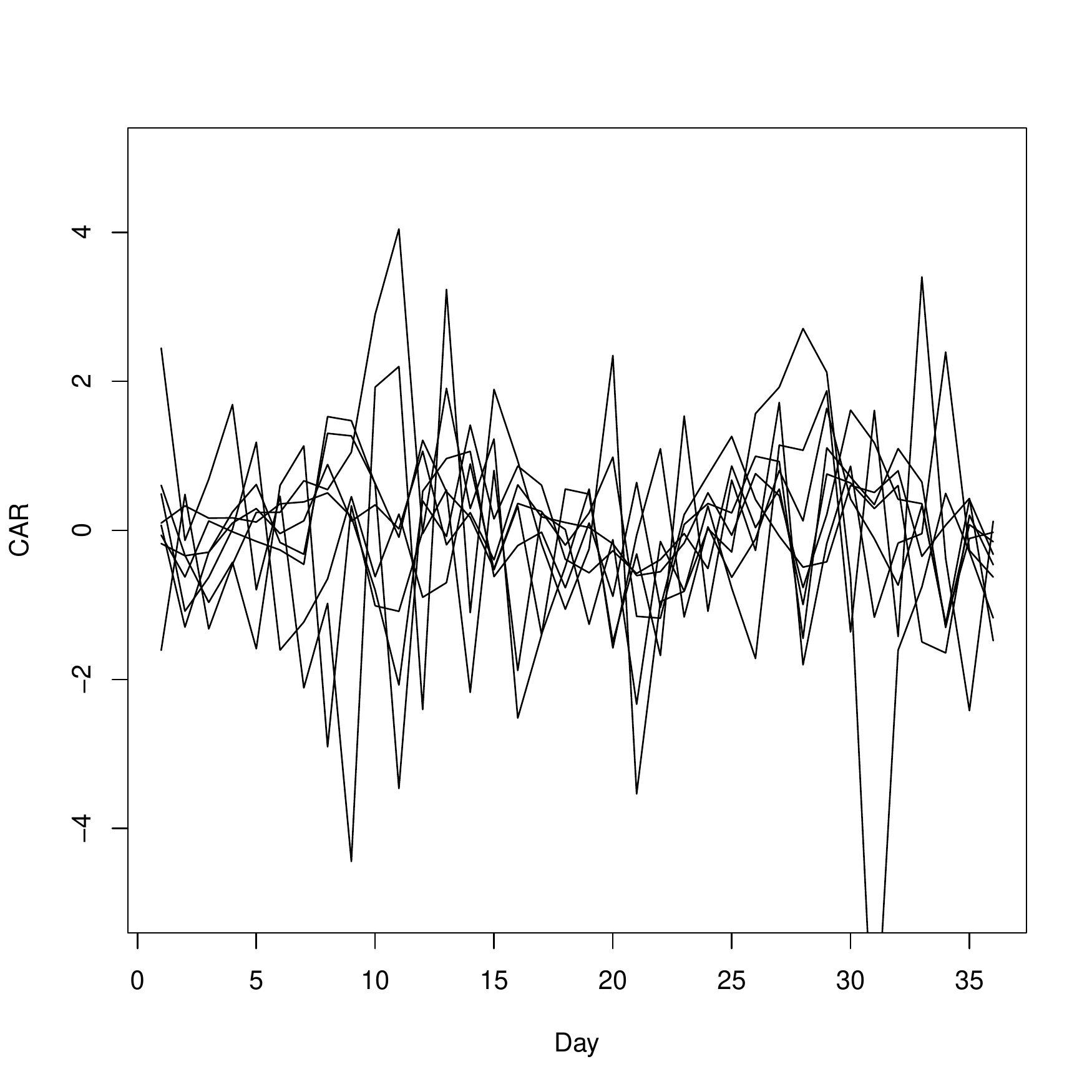} &
			\includegraphics[width=6cm,height=4cm]{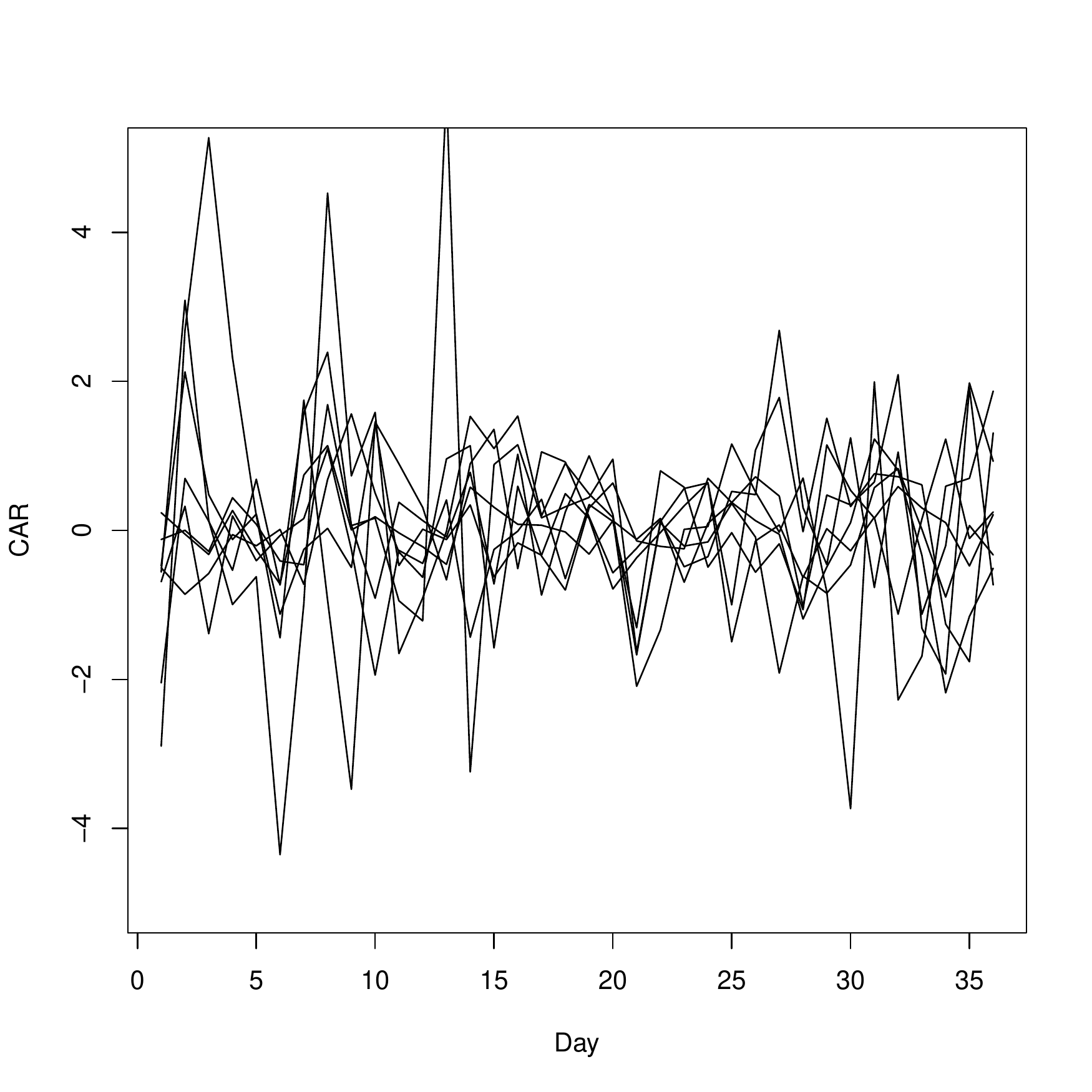} \\
			(e) Listed 2018 & (f) Delisted 2018\\
		\end{tabular}
	\end{center}
\end{figure}

Figure \ref{fig:yearcomp} provides a graphical impression of the differences across years showing the first year for which there are listings and de-listings available, 2006, the financial crisis of 2008 and the most recent set of observations in 2018. In all cases the vertical axes are plotted from -4 to 4 to enable comparability between plots. In the top row there is some evidence of movement pre-announcement in the delisting plots, whilst the CARs on stocks gaining listing only really move upward around day 18 to day 24. 2008 was much more erratic with many shares going outside the +/- 5\% range shown on the plots. De-listings were calmer with a generally positive effect seen post effective date. This contrasts with a broadly negative level of CAR for those firms gaining listing in 2008. In 2018 the pre-announcement effect is more visible as the various listed shares track each other more closely, the correction effect can also be seen in panel (e). These plots remind that amongst the average effects reported there will invariably be heterogeneities amongst stocks. Investors pursuing a particular strategy based on averages would always be reminded of this. 

Overall the set of treated firms is limited to only a few per year. Whilst further division of the dataset would be desirable to control for other known heterogeneity sources it becomes difficult to obtain a significant sample size for statistical inference. In this way an opening for more exploratory data science approaches may be found. Such extensions into data science are left beyond the scope of this paper. Here focus remains on the means through which abnormal returns are constructed, the contrast between the generalised synthetic control and CAPM. Resolving that question then makes all of the analyses that follow more credible. 

\section{Discussion}
\label{sec:discuss}

This paper seeks to understand more of the impact of firm CSR performance on their stock returns. It achieves this goal using a binary rating of whether a firm receives their listing, or not. There are many reasons to call into question such an arbitrary measure, but as outlined there is much to be said for a simple measure. Ability to interpret is critical and the binary approach does permit a clear communication with investors and the public alike. Such appeal has led to a wealth of literature capturing CSR in this binary way. Event studies become a viable method as the listing is fixed externally and is not related to the level of returns a share is experiencing at the time of the listing, or de-listing. 

Through the construction of a synthetic control potential post-evaluation period impacts on non-treated shares are accounted for in a way existing event studies have failed to do. By comparing a listed share to the performance of a portfolio of its' peers a greater understanding of the listing impact is gained. It is seen that there is an impact three weeks ahead of the effective date. Confirmation is found of a positive listing effect in the pre-change period, and a small correction in the days following the announcement. Precise durations differ because of our use of two dates, which affords our results greater accuracy to the motivational story than is found in past works with only one change date. However, in all cases the size of the abnormal return is much larger in the synthetic control. There is a definite argument for incorporating relative performance to avoid such effects being masked by linear models.

Synthetic controls can offer potential new insights for a series of treatments in finance, such as the impact of cross-listing, option availability and changes to trading rules. All of these would represent interesting applications to complement this study and the connection study of \cite{acemoglu2016value}. Here assets are used as a time-invariant control because of the comparatively large size of DJSI listed firms compared to the majority of non-DJSI listed firms. Extending the set of controls, including introducing time-variant controls, becomes increasingly possible. However, the low error within the simple fit lends a tractability to the work presented here. Likewise the approach may be fitted to intra-day data, although appropriate account for noise would be beneficial if making such an extension. This paper highlights such potential and the value of controlling for post-treatment events.

Listing on the DJSI sends an important signal to the market that a firm has achieved the highest standards of CSR. However, the effect on investors has long been considered ambiguous. Increased demand from consumers has potential to raise profitability, but in turn this delivers a stability that means lower returns are required to compensate for risk in the share price. Over and above any other impacts upon the returns of newly-listed DJSI members it is shown that abnormal returns fall when the market becomes aware of the listing. Negative effects quickly dissipate leaving an insignificant impact of DJSI listing on stock-returns. 

\section{Summary}
\label{sec:conclude}

Being listed to the DJSI sends a clear signal to the market that a firm has obtained a high level of CSR performance, and that it will be treated as such by the market. There have been numerous attempts to capture this effect but they either fail to account for important control variables, such as the two-sample approach, or they require careful matching to focus on the true change effect. By exploring the generalised synthetic control \citep{xu2017generalized} as a useful multi-treatment version of the \cite{abadie2010synthetic} method this paper has demonstrated strong abnormal returns for stocks which list on the DJSI North America. These returns far out-rated those suggested by CAPM and produced listing effects greater than those from comparing new joiners CAPM CAR with the CAR of the controls. Where two sample approaches indicate strong positive CARs between announcement and listings becoming effective and listing the synthetic control approach does not. 

We conclude that in both listing and de-listing cases there are higher returns initially, but that as the news goes public the response reverses to bring stocks back to their usual return levels. In the delisting case the generalised synthetic control results shows effects fade quicker, brining stocks back to their relative position opposite their industry competitors far quicker than in the earlier years of the DJSI North America. When contrasitng across time the difference in the financial crisis is readily apparent. Equally the time effects shown in this paper show that uncertainty about the benefits of CSR status, and hence the returns to CSR, have varied magnitude over time.  

There is scope to introduce more control variables to hone the match of the portfolio, and models beyond the CAPM could be useful. Splitting the time period may be fruitful, as the financial crisis is well known from the literature for creating an important role for socially responsible investment. Further extension could be made to winzorise the returns data, or to relax the assumption that stocks must have all of their data present. Although computationally intensive that remains an option for further work. Notwithstanding these questions the results produce cast important light on a positive benefit of listing which appears over and above the returns the treated stock would have obtained having not been listed.

\bibliography{gsynth2a}
\bibliographystyle{apalike}

\end{document}